\documentstyle[emulateapj,side,psfig]{article}

\begin{document}

\submitted{{\sc The Astrophysical Journal Supplement Series} 126: 209-269, 2000 February}
\BeforeBegin{abstract}{\centerline{\small\it{Recieved 1999 April 12; accepted 1999 August 23rd}}}
\newenvironment{inlinetable}{%
\def\@captype{table}%
\noindent\begin{minipage}{0.999\linewidth}\begin{center}\footnotesize}
{\end{center}\end{minipage}\smallskip}

% Define items
\def\et{{\it et al.\,}}
\def\etal{{\it et al.\,}}
\def\eg{{\it e.g.\,}}
\def\etc{{\it etc.\,}}
\def\ie{{\it i.e.\,}}
\def\lsim{\mathrel{\rlap{\lower 4pt \hbox{\hskip 1pt $\sim$}}\raise 1pt\hbox{$<$}}}
\def\gsim{\mathrel{\rlap{\lower 4pt \hbox{\hskip 1pt $\sim$}}\raise 1pt\hbox{$>$}}}
\def\rejected{B\,}
\def\included{A\,}
\def\DSS{C\,}
\def\masks{D\,}

\setlength\parindent{0.in}

\title{The Bright SHARC Survey: The Cluster Catalog\altaffilmark{1}}
 
\author{A. K. Romer\altaffilmark{2}, R. C. Nichol\altaffilmark{3,4}}
\affil{Dept. of Physics, Carnegie Mellon University, 5000. Forbes Ave.,
Pittsburgh, PA--15213. (romer@cmu.edu, nichol@cmu.edu)}
 
\author{B. P. Holden\altaffilmark{2,3}}
\affil{Dept. of Astronomy and Astrophysics, University of Chicago, 5640 S. 
Ellis Ave., Chicago, Il-60637. (holden@oddjob.uchicago.edu)}

\author{M. P. Ulmer \& R.A. Pildis}
\affil{Dept. of Physics and Astronomy, Northwestern University, 
2131 N. Sheridan Rd., Evanston, Il-60207. (ulmer@ossenu.astro.nwu.edu)}

\author{A. J. Merrelli\altaffilmark{2}}
\affil{Dept. of Physics, Carnegie Mellon University, 5000. Forbes Ave.,
Pittsburgh, PA--15213. (merrelli@andrew.cmu.edu)}

\author{C. Adami\altaffilmark{4,5}}
\affil{Dept. of Physics and Astronomy, Northwestern University, 
2131 N. Sheridan Rd., Evanston, Il-60207. (adami@lilith.astro.nwu.edu)}

\author{D. J. Burke\altaffilmark{3,6} \& C. A. Collins\altaffilmark{3}}
\affil{Astrophysics Research Institute, Liverpool John Moores
University, Twelve Quays House, Egerton Wharf, Birkenhead, L41 1LD, UK. 
(burke@ifa.hawaii.edu \& cac@astro.livjm.ac.uk)}

\author{A. J. Metevier} 
\affil{UCO/Lick Observatory, University of California, Santa Cruz, 
CA 95064 (anne@ucolick.org)}

\author{R. G. Kron}
\affil{Dept. of Astronomy and Astrophysics, University of Chicago, 5640 S. 
Ellis Ave., Chicago, Il-60637. (rich@oddjob.uchicago.edu)}

\author{K. Commons\altaffilmark{2}}
\affil{Dept. of Physics and Astronomy, Northwestern University, 
2131 N. Sheridan Rd., Evanston,
Il-60207. (commons@lilith.astro.nwu.edu)}

\altaffiltext{1}{Based on data taken at the European Southern Observatory,
Kitt Peak National Observatory, Cerro Tololo Interamerican 
Observatory, Canada France Hawaii \& Apache Point Observatory} 

\altaffiltext{2}{Visiting Astronomer, Kitt Peak National Observatory. 
KPNO is operated by AURA, Inc.\ under contract to the National Science
Foundation.} 

\altaffiltext{3}{Visiting Astronomer, European Southern Observatory.} 

\altaffiltext{4}{Visiting Astronomer, Canada France Hawaii Telescope.}
 
\altaffiltext{5}{Also affiliated with: Laboratoire d Astronomie 
Spatiale, Traverse du Siphon, 13012 Marseille, France.}

\altaffiltext{6}{Current address: Institute for Astronomy, Univ. of Hawii,
2680 Woodlawn Drive, Honolulu, HI 96822.}

\begin{abstract}
We present the Bright SHARC (Serendipitous High--Redshift Archival
ROSAT Cluster) Survey, which is an objective search for
serendipitously detected extended X-ray sources in 460 deep ROSAT PSPC
pointings. The Bright SHARC Survey covers an area of 178.6 deg$^2$ and
has yielded 374 extended sources. We discuss the X-ray data reduction,
the candidate selection and present results from our on--going optical
follow-up campaign. The optical follow-up concentrates on the
brightest 94 of the 374 extended sources and is now 97\% complete. We
have identified thirty-seven clusters of galaxies, for which we
present redshifts and luminosities. The clusters span a redshift range
of $0.0696<z<0.83$ and a luminosity range of $0.065<L_{\rm x}<8.3
\times 10^{44}$ erg s$^{-1}$ [0.5-2.0 keV] (assuming $H_0$ = 50 km
s$^{-1}$ Mpc$^{-1}$ and $q_0=0.5$). Twelve of the clusters have
redshifts greater than $z$=0.3, eight of which are at luminosities
brighter than $L_{\rm x}=3\times 10^{44}$ erg s$^{-1}$. Seventeen of
the 37 optically confirmed Bright SHARC clusters have not been listed
in any previously published catalog. We also report the discovery of
three candidate ``fossil groups'' of the kind proposed by
\cite{ponman94b}.

\end{abstract}
\subjectheadings{catalogs --- galaxies: clusters: general --- galaxies: distances and redshifts --- surveys --- X-rays: galaxies}

\section{Introduction}

Clusters of galaxies play a key role in constraining cosmological
models.  It has been shown (\eg \cite{Oukbir}; \cite{Carlberg};
\cite{henry97}; \cite{Bartlett}; \cite{viana98}) that measurements of
the cluster number density, and its evolution, play an important role
in the derivation of the mean mass density of the Universe,
$\Omega_m$. At present, there is a large dispersion in the values of
$\Omega_m$ derived from measurements of the cluster number density;
\eg $\Omega_m=0.2^{+0.3}_{-0.1}$ (\cite{Bahcall}),
$\Omega_m=0.4^{+0.3}_{-0.2}$ (\cite{Borgani}), $\Omega_m=0.5^{\pm
0.14}$ (\cite{henry97}), $\Omega_m=0.85^{\pm0.2}$ (\cite{Bartlett}),
$\Omega_m=0.96^{+0.36}_{-0.32}$ (\cite{Reichart1}).

To fully exploit clusters as cosmological tools one needs to have
access to large, objectively selected, cluster catalogs which cover a
wide redshift range. Most cluster catalogs constructed prior to 1990
had a very limited redshift range and were not constructed in an
objective manner (\eg \cite{abell56}; \cite{abell89}). However, recent
developments, such as CCD mosaic cameras, optical plate digitizers and
imaging X-ray satellites, have resulted in a growing number of high
quality cluster catalogs. These include optically selected cluster
samples derived from digitized plate material, \eg the EDCC
(\cite{lumsden92}) and the APM (\cite{Dalton}), or from CCD imaging
surveys, \eg the PDCS (\cite{post96}) \& the ESO Imaging Survey
(\cite{DaCosta}).  X-ray selected cluster samples derived from imaging
X-ray satellite data include those from the Einstein mission, \eg the
EMSS cluster sample (\cite{gioia90}), and those from the ROSAT
(\cite{Truemper}) mission.

The various ROSAT cluster catalogs divide into two categories; those based on
ROSAT All-Sky Survey (RASS) data and those based on ROSAT pointing data.  The
former category includes the SRCS (\cite{Romer1}), XBACS (\cite{Ebeling1}),
BCS (\cite{Ebeling2}), REFLEX (\cite{Bohringer}; \cite{DeGrandib}) and NEP
(\cite{gioia99}; \cite{henryet97}) surveys.  Examples of surveys based on ROSAT
pointing data are the SHARC (\cite{Collins}); RIXOS (\cite{Castander}); RDCS
(\cite{Rosati}); WARPS (\cite{Jones}) and 160deg$^2$ (\cite{Viklinin1})
surveys.  The ROSAT instrument of choice for cluster surveys has been the
PSPC, which combines imaging capabilities with a large field of view
(2$^{\circ}$ in diameter), low background contamination and some spectral
resolution. The angular resolution of the ROSAT PSPC is better than that of
Einstein, allowing one to take advantage of the extended nature of cluster
emission to distinguish clusters from X-ray point sources, \eg AGN and
quasars.  Moreover, the enhanced sensitivity of ROSAT over Einstein means that
ROSAT cluster surveys can reach fainter flux limits than the EMSS.

The RASS surveys have yielded several important insights into the
clustering properties (\cite{romer94}) and evolution
(\cite{Ebeling97}; \cite{DeGrandia}) of the $z<0.3$ cluster
population. At higher redshifts, the ROSAT pointing data surveys have
shown that there is no evidence for evolution in the cluster
population at luminosities fainter than $L_{\rm x}=5\times10^{44}$ erg
s$^{-1}$ [0.5-2.0 keV] and redshifts less than $z\simeq 0.7$
(\cite{Nichol97}; \cite{Burke}; \cite{Collins}; \cite{Viklinin2}; 
\cite{Rosati}; \cite{Jones}; \cite{Nichol99}). 
At brighter luminosities, the 160deg$^2$ and Bright SHARC surveys,
have provided evidence for negative evolution (\cite{Nichol99};
\cite{Viklinin2}) similar to that seen in the EMSS cluster sample
(\cite{henry92}; \cite{Reichart1}).

The SHARC (Serendipitous High--Redshift Archival ROSAT Cluster) survey
was designed to optimize studies of X-ray cluster evolution and
combines two complementary surveys; a narrow area deep survey and a
wide area shallow survey. The former, known as the Southern SHARC, has
been described elsewhere, (\cite{Collins}; \cite{Burke}). We introduce
the latter survey, the Bright SHARC, here. Unlike the Southern SHARC,
the philosophy of the Bright SHARC has been to achieve maximum areal
coverage rather than maximum sensitivity. The Bright SHARC survey
covers a total area of 178.6 deg$^2$ and has yielded a catalog of 37
clusters with fluxes $\geq$1.63$\times 10^{-13}$ erg s$^{-1}$
cm$^{-2}$. In contrast, the Southern SHARC survey covers only 17.7
deg$^2$, but has yielded a similar number of clusters (36) with fluxes
$\geq$4.66$\times 10^{-14}$ erg s$^{-1}$ cm$^{-2}$.

We describe below the reduction of the 460 ROSAT PSPC pointings in the
Bright SHARC survey (\S\ref{pspcdata}), our source detection
methodology (\S\ref{sourcedetection}) and the selection and optical
follow-up of cluster candidates (\S\ref{extent} \& \S\ref{identify}).
In sections \S\ref{cluster_section} and \S\ref{results} we present and
discuss the Bright SHARC cluster catalog. Throughout this paper we use
$H_0$ = 50 km s$^{-1}$ Mpc$^{-1}$ and $q_0=0.5$ and define $f_{-13}$
and $L_{44}$ to be the unabsorbed flux (observer frame) and luminosity
(rest frame) in the [0.5-2.0 keV] energy band in units of 10$^{-13}$
erg s$^{-1}$ cm$^{-2}$ and 10$^{44}$ erg s$^{-1}$, respectively.

\section{Transfer and Reduction of ROSAT PSPC Pointing Data}
\label{pspcdata}

The pointed PSPC data used in the construction of the Bright SHARC was
obtained from the HEASARC ROSAT data archive using an automated FTP
process over a period of two years starting in June 1995. The criteria
used to select data from the archive were: {\it i}) a listed exposure
time greater than 10ksecs and {\it ii}) an absolute Galactic latitude
greater than 20 degrees.  Based on these criteria, 638 PSPC pointings
were transferred to local machines and then reduced using a pipeline
processing based on the Extended Source Analysis Software (ESAS)
package (\cite{Snowden}).

An overview of our pipeline is as follows. First, the raw data were
sorted into good and bad time intervals. Bad intervals were defined as
those in which the background level was higher than 170 counts
s$^{-1}$.  Data obtained during bad intervals were discarded. The
remaining data were binned, as a function of position, into seven
different energy bands --which \cite{Snowdenb} define as R1 through
R7-- to produce seven $512\times512$ pixel maps with a pixel scale of
$14''.947$.  The 5 highest energy bands, R3 through R7, were then
co-added to produce a hard band [0.4-2.0 keV] count rate map for each
pointing. Accompanying each count rate map was a count rate
uncertainty map and a vignetting corrected exposure map.

For each of the 638 pointings reduced by the ESAS pipeline, the ROSAT pointing
name (column 1), the J2000 position (columns 2 \& 3), Galactic latitude in
degrees ($b$, column 4), the exposure time in seconds (column 5) and pointing
target (column 6) are listed in appendices \included \& \rejected.  Appendix
\included lists the 460 pointings selected to form the Bright SHARC
survey. Two points should be noted about these 460 pointings.  First, 371 of
the pointings, those with 3 character extensions \eg `n00', `a01' \etc, were
processed after an important change was made to the Standard Analysis Software
System (SASS). This change effected those pointings for which the total
exposure time was broken up into more than one observation interval. After the
change, each observation interval was analyzed separately, whereas before the
change they were analyzed together.  Of these 371 pointings, only 45 have more
than one observing interval. For simplicity we decided to include only the
longest observing interval in our analysis. Second, the Bright SHARC survey
does not include the central $2.'5$ radius region of PSPC (see \S
\ref{extent}). This means that we were able to include several pointings in
the survey which had intrinsically extended central targets, \eg galaxies and
clusters -- as long as those targets did not extend beyond $2.'5$. We discuss
cases where Bright SHARC clusters are detected in pointings with cluster
targets in Table \ref{clusters} (see \S\ref{cluster_section}) and
\S\ref{contam}.

The 178 pointings listed in appendix \rejected were not included in
the Bright SHARC because either an extended X-ray (or optical) source
covers most of the field of view, or the pointing is within $<
6^{\circ}$ of the Magellanic clouds.  Extended X-ray and optical
objects include low redshift Abell clusters and Galactic globular
clusters. The pointings listed in Appendix \rejected were removed
after visual inspections of the reduced X-ray data and the Digitized
Sky Survey. Despite its subjective nature, this procedure does not
undermine the serendipitous nature of the SHARC survey, since it was
performed before the cluster candidate list was constructed. We have
indicated in column 7 of appendix \rejected why each pointing was
rejected from the survey.
 
\section{Source Detection}
\label{sourcedetection}

Our source detection algorithm was based on wavelet--transforms
(\cite{Slezak}).  For our purposes, we required a detection algorithm
which ({\it i}) was sensitive to both extended and point--like
sources, ({\it ii}) worked in crowded fields and ({\it iii}) took into
account a varying background level.  Moreover, we wanted our method to
be as simple as possible, so that we could define our selection
function {\it a posteriori} using simulations.  With these concerns in
mind, we chose to convolve the PSPC count rate maps with a spherically
symmetric, ``Mexican-hat'' wavelet. This wavelet, in one dimension, is
given by:

\begin{equation}
w(x)=(1-\frac{x^2}{a^2})\,{\huge\it e}^{-\frac{x^2}{2a^2}},
\label{waveleteqn}
\end{equation}

and is the second derivative of a Gaussian (\cite{Slezak}) of width
$\sigma=a$.  The radially averaged point-spread function of the ROSAT
PSPC can be approximated to a Gaussian (Hasinger \et 1992), so this
wavelet is well suited to the detection of sources in PSPC images.  An
additional attraction of this wavelet is that it can be used to
determine the extent of a source, since it has a width of $2\times a$
at its zero--crossing points. A wavelet transform of a PSPC count rate
map will, therefore, produce a wavelet coefficient map in which all
the sources are bounded by a ring of zero values. The diameter of
these zero crossing rings provides a direct measure of the source's
extent in wavelet space.

Ideally, ``$a$'' should be scaled logarithmically to provide
statistically independent wavelet images over the whole range of real
and k--space. For any given source, the wavelet coefficient will have
a maximum when the value of ``$a$'' matches the sigma of the best fit
Gaussian. However, the use of multiple wavelets would make {\it a
posteriori} simulations of the selection function very complex and CPU
intensive.  We therefore decided to use only one wavelet convolution
($a=3$ pixels or 45$''$) in our source detection pipeline. This
particular wavelet was found, empirically, to be the best compromise
between smaller wavelets, which tended to fragment extended sources,
and larger wavelets, which tended to blend neighboring sources.  The
penalty for this simplification was the inclusion of some blended
sources in our extended candidate list (\S\ref{identify}) and
underestimated cluster count rates (\S\ref{aperture}).

Sources were identified in the wavelet coefficient map by selecting
pixels with coefficients above a given threshold. This threshold was
set, empirically, to be 7 sigma above the peak of the coefficient
distribution.  The thresholding technique only highlights the cores of
each source, since that is where the wavelet coefficients are highest,
so a ``friends--of--friends'' analysis was run to identify other
associated pixels. This was done by growing the sources outwards in
the wavelet coefficient map until they reached the zero crossing ring.
Once the source boundaries were defined, best fit centroids and
ellipses were computed. A filling factor was also derived for each
source. This was defined as ratio of the area within the fitted
ellipse to the area within the zero--crossing boundary.  For $f=1$,
the ellipse fits the source shape exactly, while $f>>1$ indicates the
presence of blended sources (dumb-bell shapes) or percolation runaways
(filamentary shapes).

In total, 10,277 sources were detected in the 460 pointings. To keep
track of all of these sources, and their boundaries, a mask file was
generated for each pointing so that pixels associated with sources
could be distinguished from those that were not. For each source, a
$51\times51$ pixel box, with the source at its center, was extracted
from the count rate map.  An average background for the box was
calculated using all pixels not flagged as belonging to sources. The
count rate for the central source was then derived by subtracting this
background (appropriately scaled) from the sum of the pixels enclosed
by the source boundary. We used this method because it was easy to
apply to the thousands of sources detected in the Bright SHARC
Survey. (We will refer to the count rates derived in this manner as
``wavelet count rates'', $cr_W$, hereafter.)  However, the method has
the disadvantage of underestimating the true count rate if the source
is extended beyond the wavelet boundary (in \S\ref{aperture} we
describe an alternative method used to derive count rates for known
clusters). An approximate signal-to-noise value for each source was
also calculated using the count rate uncertainty maps produced by
ESAS.

It should be noted that certain pixels, those which received less than
half the exposure time of the central pixel in the count rate maps, were not 
included in the ``friends--of--friends'' analysis. Such pixels 
included those in the shadow of the PSPC window support structure and those
at the edge of the field of view. These regions, which are noisier than
those that were well exposed, were not used to define source centers, 
shapes or wavelet count rates. 

\section{Selection of Cluster Candidates} 
\label{extent}

The majority of X-ray sources can be considered point like in their
spatial properties, \eg stars and AGN. In the minority are objects
with complex and extended X-ray profiles, such as supernova remnants,
galaxies and clusters of galaxies. Of these, only clusters are large
enough and bright enough to be detected as extended beyond $z\simeq
0.1$. Therefore the strategy adopted by the SHARC has been to search
for clusters only among those ROSAT sources that have a significant
extent. This reduces the required optical follow-up significantly. The
disadvantage of this approach, however, is that some clusters, \eg
those with compact surface brightness distributions, may be excluded
from the survey.

Bright SHARC Cluster candidates were selected from the 10,277 sources
found in the survey using the following six criteria: The source had
to ({\it i}) have a signal-to-noise ratio greater than 8, ({\it ii})
its centroid had to fall within 90 pixels ($22'.4$) of the pointing
center, ({\it iii}) its centroid had to fall more than 10 pixels
($2'.5$) from the pointing center, ({\it iv}) its filling factor had
to be less than $f=1.3$, ({\it v}) it had to be more than $3\sigma$
extended and ({\it vi}) it have to have a count rate higher than
0.01163 counts s$^{-1}$. The imposition of these criteria cut down the
source list from 10,277 (total) to 3,334 (criterion {\it i}) to 1,706
(criterion {\it ii}) to 374 (criteria {\it iii} to {\it v}) to 94
(criterion {\it vi}). Criterion ({\it i}) was imposed because it has
been shown (Wirth \& Bershady, in preparation) that extent measures can
only be derived with confidence for sources meeting a minimum
signal-to-noise threshold.  Criterion ({\it iii}) was applied to avoid
including the intended target of the pointing in the candidate
list. Criterion ({\it iv}) was set empirically with the aim of
reducing the number of blended sources and percolation runaways in the
candidate list. The rationale for the other criteria is provided
below.

\begin{figure}
\centerline{\psfig{figure=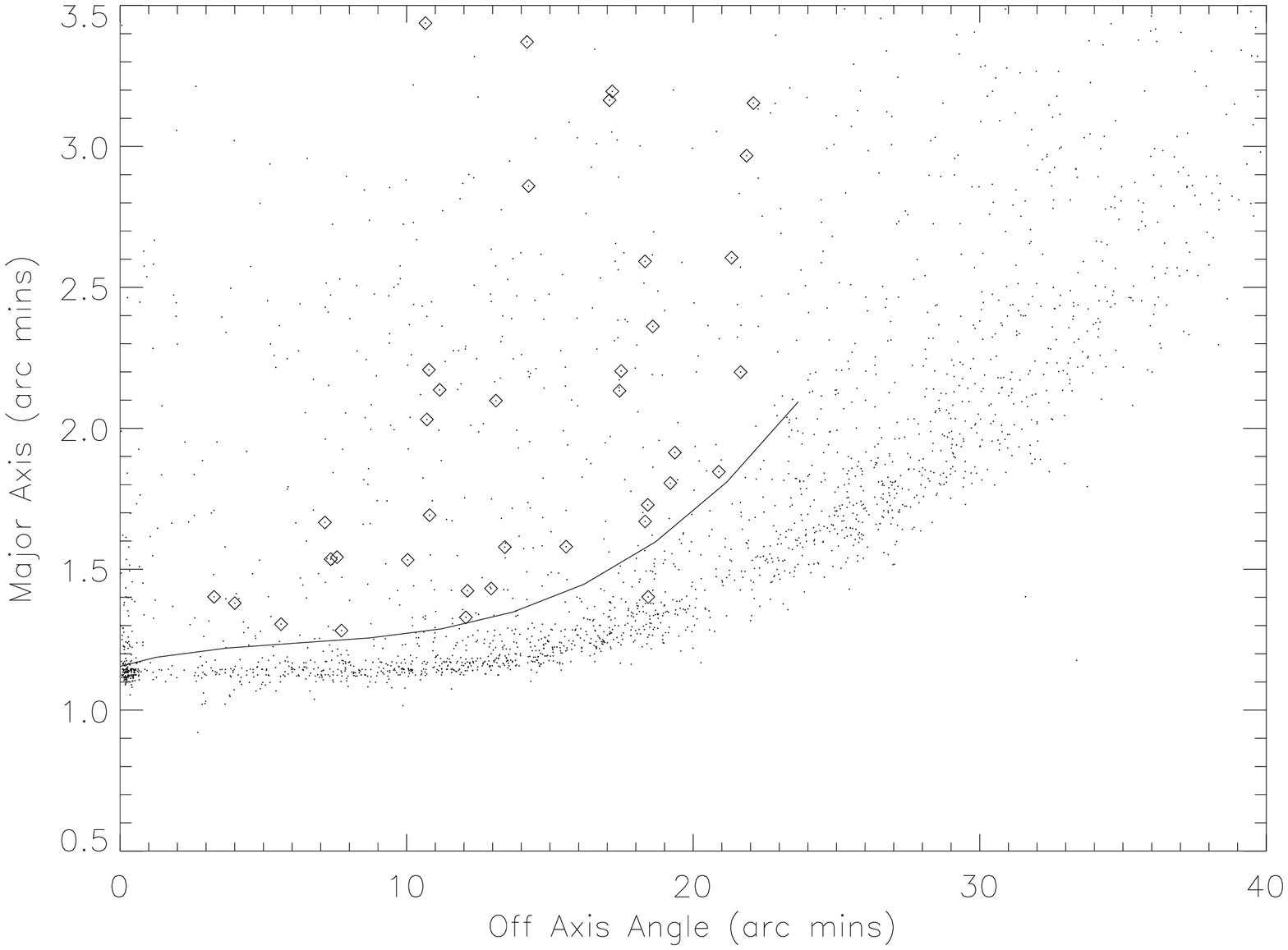,height=3.5in,angle=90}}
\centerline{\psfig{figure=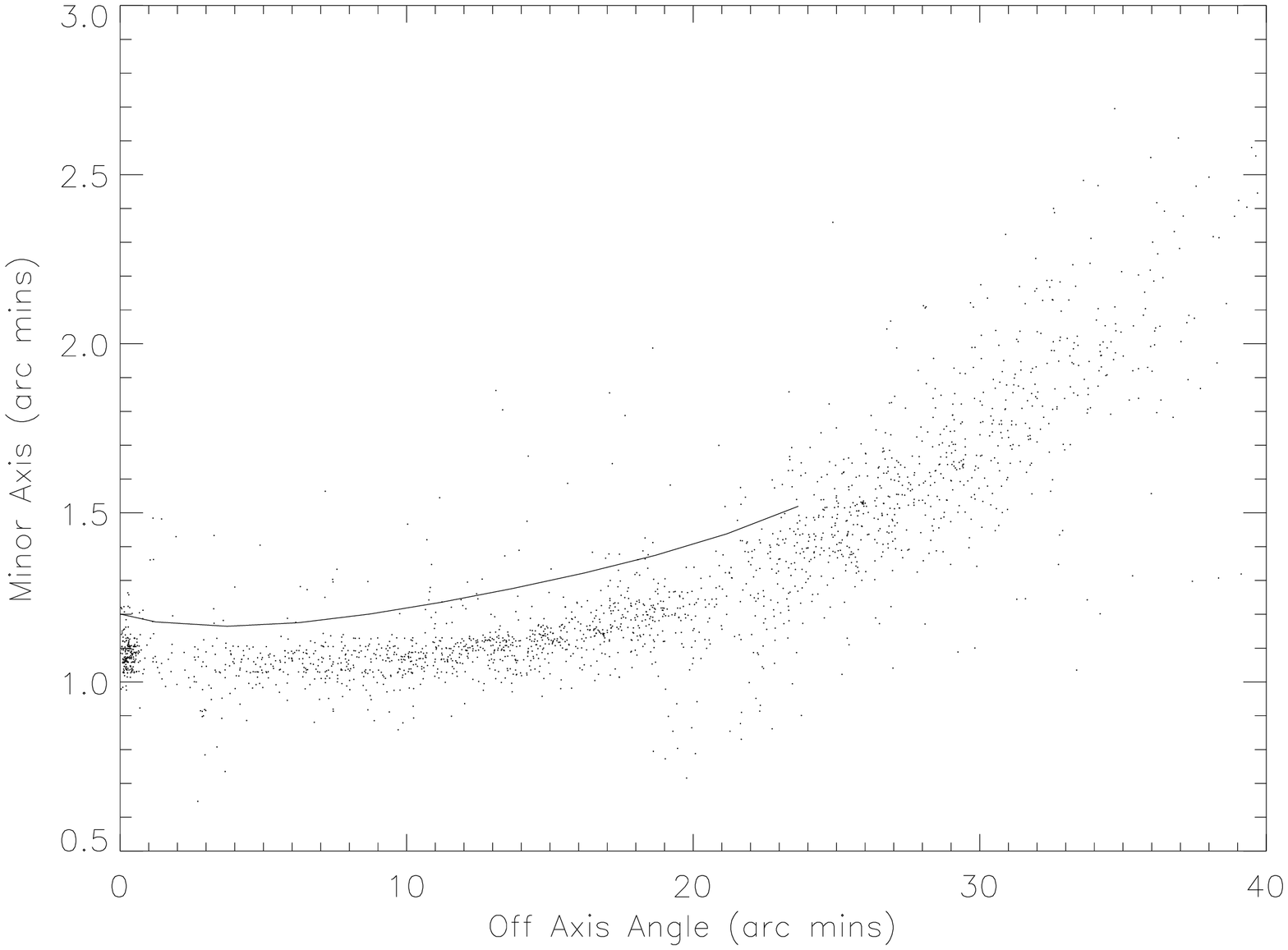,height=3.5in,angle=90}}
\caption{The distribution of major and minor axes for the 3,334
S/N$>$8 sources in the Bright SHARC survey as a function of off-axis
angle.  The solid lines correspond to the fitted three-sigma extent curves; 
any sources falling above these lines are classified as extended. 
For illustration purposes, we have plotted, as open diamonds, the points 
corresponding to the thirty-seven Bright SHARC clusters on the major axis plot 
(the numerical values for these points can be found in Table \ref{clusters}).}
\label{extentcurves}
\end{figure}

The point-spread function of the PSPC degrades significantly as one
moves out from the center of the detector (\cite{Hasinger}).  It
therefore becomes increasingly difficult to distinguish extended
sources from point sources as the off-axis angle increases. To
overcome this, we used all 3334 of the S/N$>$8 sources in our survey
to study statistically how source size varies as a function of
position on the PSPC. The method used has been described previously
(\cite{Nichol97}), but we include an overview here for completeness.
Figure \ref{extentcurves} shows the distribution of source size (as
defined by the lengths of the major and minor axes of the best fit
ellipses) as a function of off-axis angle.  After collecting these
data into 10 pixel bins, we were able to determine how the mean and
FWHM of the distribution varied with off-axis angle. (Beyond an
off-axis angle of 90 pixels, the dispersion in source sizes became too
large to define a reliable FWHM, hence the imposition of criterion
{\it ii}). Under the assumption of a Gaussian distribution, the FWHM
values were converted into sigma values and a three sigma curve was
determined by fitting a 4th order polynomial to the [mean$+3\sigma$]
values. A source was defined to be extended if it had a major and/or a
minor axis more than $3\sigma$ from the mean.

In total, 374 sources were found to meet criteria ({\it i}) through
({\it v}).  These are listed in Appendix E in right ascension
order. Wavelet countrates ($cr_W$) are given for each source in units
ot $1\times10^{-2}$ counts s$^{-1}$ (column 4).  We note that
duplicate entries, \eg RX J0056.5--2730 -- which was detected in two
pointings, wp700528 and rp701223n00 -- have not been excised from this
list.  The fluxes for these 374 sources (assuming thermal spectra see
\S\ref{convert}) range from $0.2\lsim f_{-13}\lsim 40$. In the
interests of completing the optical follow-up in a timely fashion, it
was decided to concentrate only on the brightest of these 374. An
arbitrary count rate cut ($cr_W>0.01163$) was imposed to reduce the
sample size to roughly 100 (criterion {\it vi}). At the redshift of
the most distant cluster in the EMSS sample ($z$=0.81), this count
rate corresponds to a luminosity of $\simeq3.9L_{44}$, which is
approximately equal to locally determined values of $L_{\star}$, \eg
$L_{\star}$=5.7$L_{44}$ (\cite{Ebeling97}), $L_{\star}$=3.8$L_{44}$
(\cite{DeGrandib}).

The total areal coverage of the Bright SHARC survey is $178.6$ square
degrees. This value was determined by calculating the area available
for candidate detection in each of the 460 pointings in the survey.
This area includes all pixels at radii greater than $2'.5$ and less
than $22'.4$ which ({\it i}) had exposure times more than half that of
the central pixel and ({\it ii}) did not overlap pixels in a higher
exposure pointing. (There were 21 pairs of pointings with some overlap
between them.)

\section{Identification of Extended Sources}
\label{identify}

We present the 94 unique\footnote{Duplicate entries for RX
J0237.9--5224 and RX J1211.2+3911 have been removed.} extended sources
in the Bright SHARC survey in Table \ref{XrayBrightSHARC} and Appendix
\DSS.  For each candidate, we provide the source name (column 1), its
J2000 position (columns 2 \& 3), the wavelet count rate [0.4-2.0 keV]
($cr_W$ in units ot $1\times10^{-2}$ counts s$^{-1}$ column 4), the
pointing in which it was detected (column 5), the source type (column
6), and the method used to identify the source (column 7). Alternate
source names and redshifts (where available) are listed in column
8. We note that Abell clusters (\cite{abell56}; \cite{abell89}) are
denoted by `A'. Likewise for EMSS sources (`MS', \cite{stocke}), 160
deg$^2$ clusters (`V', \cite{Viklinin1}, V98 hereafter), Hickson
groups (`HCG', \cite{hickson82}) and Zwicky clusters (`Z',
\cite{zwicky}).  When an object is listed in more than one catalog, we
have defaulted to the name given in the older catalog, \eg for RX
J0237.9 we have listed the Abell number (A3038), not the V98 number (V28).

In Appendix \DSS, we present small ($6'.6\times6'.6$) Digitized Sky
Survey (DSS) images of each of the $94$ extended sources listed in
Table \ref{XrayBrightSHARC}.  The source outlines, as defined by our
friends-of-friends analysis are overlaid on these images. We note that
the source centroids were defined in a weighted fashion and do not
necessarily coincide with the geometric center of the source
outline. No external astrometric solution was applied before making
these DSS images, because the expected pointing offset is much smaller
($\lsim 6''$) than the typical size of one of our extended sources.

In some cases it was possible to identify the source using the DSS
images alone. For example, the X-ray emission from source RX
J0324.6-5103 is clearly associated with a bright star (HD21360). This
source was flagged as extended because emission from the star was
blended by the friends-of-friends analysis with the (fainter) emission
from a neighboring point source. (The X-ray surface brightness
contours for this source show a secondary peak centered on the faint
DSS object to the lower left of the source outline.)  In other cases,
the source outline, and/or the surface brightness contours, are
indicative of blended emission but no obvious counterpart could be
found on the DSS images, \eg RX J0947.8+0741.  When the DSS (or X-ray)
images played a role in the identification of a source, a `D' (or `X')
is listed in column 7 of Table \ref{XrayBrightSHARC}.

A search of the NASA/IPAC Extragalactic Database (NED) has also
provided useful information for several of the Bright SHARC extended
sources, including some cluster redshifts \eg for RX J1204.0+2807
(MS1201.5, \cite{gioia90}).  When NED yielded information was used
during the source identification, an `N' is listed in column 7 of
Table \ref{XrayBrightSHARC}.

Optical follow-up of Bright SHARC extended sources has been carried
out at a number of telescopes; the 3.5m ARC telescope at Apache Point
Observatory, the Danish 1.5-m and 3.6-m telescopes at the ESO Southern
Observatories, the 1.5-m telescope at the Cerro Tololo Inter-American
Observatory, the 3.6-m Canada France Hawaii Telescope on Mauna Kea and
the 4-m Mayall telescope at Kitt Peak National Observatory. Optical
follow-up includes CCD imaging, long slit spectroscopy and
multi-object spectroscopy. Of the 94 extended sources, to date 57 have
CCD images and 51 have been the target of spectroscopic follow-up. A
`C' in column 7 of Table\ref{XrayBrightSHARC} indicates that a CCD
image is available, whereas an `S' indicates spectroscopic follow up
by the SHARC collaboration and an `O' indicates that spectroscopy came
from private communications. To date, 91 of the 94 Bright SHARC
extended sources have been identified; 37 clusters, 41 blends, 9
galaxies and 3 galaxy groups. The symbols '+ ?' in column 8 indicate
that the identification of one of the components of a blended source
is unknown. We note that the distinctions between galaxies and groups
(see \S\ref{fossil}), and between groups and clusters, are not
absolute at the low luminosity end. For the 12 extended objects (9
galaxies and 3 groups) at redshifts less than $z=0.07$, we based our
classifications on the information provided by NED.

\section{The Bright SHARC Cluster Sample}
\label{cluster_section}

The thirty-seven clusters in the Bright SHARC are listed in Table
\ref{clusters}.  For each cluster, we list the source number (column
1), the cluster redshift (column 2), the hydrogen column density (in
units of 1$\times10^{20}$ cm$^{-2}$, column 3), the major and minor
axes (in units of $14.''947$ pixels, columns 4 \& 5), the offaxis
angle of the cluster centroid (in units of $14.''947$ pixels, column
6), the wavelet ($cr_W$) and total ($cr_T$) count rates [0.4-2.0 keV]
(in units of $1\times10^{-2}$ counts s$^{-1}$, columns 7 \& 8), the
percentage error on $cr_T$ ($\delta cr_T$, column 9), the aperture
containing 80\% of the flux from a model cluster profile ($r_{80}$,
column 10, see \S\ref{aperture}), the fraction of that aperture used
to measure the cluster count rate ($f_{80}$, column 11), the total
flux [0.5-2.0 keV] ($f_{-13}$, in units of 1$\times10^{-13}$ erg
s$^{-1}$ cm$^{-2}$, column 12), the corresponding luminosity
($L_{44}$, in units of 1$\times10^{44}$ erg s$^{-1}$, column 13), and
the temperature used to derive the flux and luminosity (T, in units of
keV, column [14]). Various notes, including alternative cluster names
and pointers to the information on the pointing target (if that target
was a cluster) are given in column (15).

The redshift distribution ({\it \={z}}=0.266) for the 37 Bright SHARC
clusters is shown in Figure \ref{zhisto}. The highest redshift, and
most luminous, cluster in the sample is RX J0152.7 ($z$=0.83). The
lowest redshift cluster in the sample is RX J0321.9
($z$=0.0696). Twenty-one of the redshifts in Table \ref{clusters} are
presented here for the first time.  These 21 include 17 clusters which
have not been listed before in any published catalog and 4 clusters
from the 160 deg$^2$ survey (V98).  We describe below how the count
rates (\S\ref{aperture}) and fluxes/luminosities (\S\ref{convert})
were derived.
 
\subsection{Total Cluster Count rate Derivation}
\label{aperture}

The method described in \S \ref{extent}, to measure wavelet count
rates for all 10,277 sources in the Bright SHARC survey, was adopted
because it was easy to apply to large numbers of sources. However, the
method is not optimal for measuring cluster fluxes. This is because no
correction is made for cluster flux falling outside the zero crossing
boundary. Moreover, when a portion of a cluster overlaps a masked out
region (\eg regions in the shadow of the support struts), the flux
from that region will not be included in the count rate.  Therefore,
for the 37 sources identified with clusters of galaxies, we have
performed a second count rate determination based on the method
adopted by \cite{holdenb}. For each of the clusters, we derived an
aperture for the flux measurement using a cluster model based on a
modified isothermal sphere:
\begin{equation}
I=\frac{I_0}{[1+(r/r_c)^2]^{3\beta -1/2}} \,\,,
\label{betamodel}
\end{equation}
where $I$ is the surface brightness at radius $r$. We used values for
the slope ($\beta=0.67$) and core radius ($r_c=250$ kpc) which are
typical for rich clusters (\cite{JonesC}) and then converted the model
from physical units to angular units using the cluster redshift. The
cluster models were convolved with the appropriate off-axis PSPC PSF
(\cite{Nichol94}) so that the radius of a circular aperture, $r_{80}$,
which contained $\simeq$80\% of the total model flux could be defined
(for $\beta=\frac{2}{3}$, $r_{80}=\sqrt{24}r_c$).  The choice of
$r_{80}$ for the aperture represents a compromise between including a
high fraction of the cluster flux and keeping down the number of
contaminating sources within the region.

The 80\% radii, $r_{80}$, are listed in column 10 of Table
\ref{clusters}, in units of $14''.947$ pixels. Since these radii could
be quite large, up to 40 pixels, some of them overlapped other
sources. If any $r<r_{80}$ pixels lay within the wavelet-defined
boundary of another source, they were masked out from the cluster
aperture. Also masked were any pixels that received less than half the
exposure time of the central pixel in the count rate map. By reference
to the cluster model, it was possible to correct for the fraction of
cluster flux lying in these masked regions. In column 11 of Table
\ref{clusters}, we list the fraction of the 80\% aperture available
for flux determination, $f_{80}$. The raw aperture count rates for
each of the 37 clusters were measured by summing the flux in the
unmasked $r<r_{80}$ pixels.

The corresponding background count rates were measured inside
$120\times120$ pixel boxes centered on the cluster position. The
background levels were measured in annuli with minimum radii of
$1\times r_{80}$ and maximum radii $3\times r_{80}$.  If these annuli
overlapped any source boundaries, any low exposure pixels, or the
edges of the $120\times120$ pixel box, the pixels in those regions
were excluded from the background calculations. In Appendix \masks we
illustrate the masked out regions for the source and background
apertures for each of the 37 Bright SHARC clusters. After subtraction
of the appropriately scaled background, the total cluster count rates
were derived by dividing by ($0.8\times f_{80}$). The background
subtracted, aperture corrected, total cluster count rates ($cr_T$) are
listed in column 8 of Table \ref{clusters}.  The one sigma errors on
the total cluster count rates are listed in column 9 of Table
\ref{clusters}. These errors were calculated by adding in quadrature
the counting errors on the cluster count rates and the background
count rates. We draw attention to three SHARC clusters with
anomalously high ($>15\%$) count rates errors; RX J0250.0, RX J1524.6,
RX J1222.1.  These clusters have much lower signal to noise values
inside the $cr_T$ apertures than in the $cr_W$ apertures,
demonstrating that the adopted cluster model (equation\ref{betamodel})
significantly over estimates the size of the aperture which encircles
80\% of the source flux.  The count rate errors are quoted as
percentages since, in the absence of systematic errors in the count
rate to flux/luminosity conversions (\S \ref{convert}), they should
also reflect the percentage errors on the flux ($f_{-13}$, column 12)
and luminosity ($L_{44}$, column 13).

\begin{figure}
\centerline{\psfig{figure=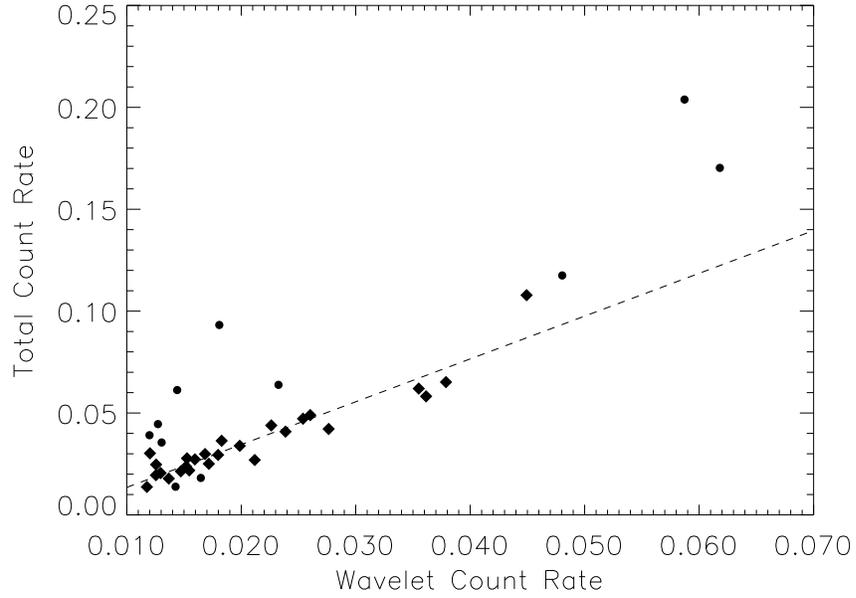,height=3.5in,angle=90}}
\caption{The wavelet count rate versus the total count rate for each
of the thirty-seven clusters in the Bright SHARC sample. The low redshift
($z<$0.15) clusters are indicated by circles. A least squares fit to
the $z>0.15$ clusters (diamonds) is shown by the dotted line (slope=2.1).}
\label{ctrcomp} 
\end{figure}

In Figure \ref{ctrcomp} we compare the initial wavelet count rates,
$cr_W$ (column 7), to the total aperture corrected count rates,
$cr_T$.  It can be seen that, as expected, the total count rate is
systematically higher than the wavelet count rate. A least squares fit
to the clusters at redshifts greater than $z$=0.15 shows that the
total count rate is typically 2.1 times higher than the wavelet
value. At lower redshifts, the correction is higher because the
clusters are significantly more extended than the $\sigma=3$ pixel
wavelet we used for source detection.  It is encouraging that the
wavelet count rate appears to be an unbiased measure of the total
cluster count rate, since we have used the $cr_W$ values to define the
count rate limit of the Bright SHARC Survey.

\subsection{Luminosity Derivation}
\label{convert}
We used the $cr_T$ count rates listed in column 8 of
Table\ref{clusters} to determine fluxes and luminosities for each
cluster.  We note that we chose to present the fluxes and luminosities
in the [0.5-2.0 keV] band, rather than in the Bright SHARC count rate
band [0.4-2.0 keV], to allow easier comparison with other
studies. Since the ROSAT PSPC provides only limited spectral
resolution, we had to assume a spectral model for each cluster to make
the conversion between measured cluster count rate and unabsorbed
flux. As is typical in X-ray cluster analyses, we adopted an emission
spectrum from hot, diffuse gas based on the model calculations of
Raymond and Smith (\cite{Raymond}).  The integrated emission from a
Raymond-Smith spectrum in the SHARC energy band (observer's rest
frame) depends on several factors; the metallicity and temperature of
the gas, the redshift of the cluster, and the absorption column along
the line of sight.  This means that the conversion between measured
aperture count rate and cluster luminosity is non trivial and must
take into account the specific properties of each cluster.  We note
that, in most cases, the dominant source of error in the derived
luminosities comes from the count rate uncertainty, which rises to
30\% in the case of RX J1524.6. However, for those clusters with well
determined count rates (30 clusters have count rate errors of less
than 10\%) it is worth making the extra effort to reduce the
systematic errors in the conversion between count rates, fluxes and
luminosities.

We have constructed a matrix of count rate to flux conversion factors
as a function of temperature, redshift and absorbing column.  (A
single, canonical, value for the metallicity -- one third the Solar
value -- was used throughout.) The conversion factors were derived
using the {\tt fakeit} command in XSPEC (version 10.00,
\cite{ArnaudK}) together with the appropriate ROSAT PSPC response
function.  Photo-electric absorption was included via the XSPEC {\tt
wabs} model, which is based on cross sections presented in
\cite{nHb}. The neutral Hydrogen column densities adopted for each
cluster are listed in column 3 of Table \ref{clusters}.  These values
were derived using the AT\&T Bell Laboratories 21 cm survey
(\cite{stark}), for clusters north of $-40^{\circ}$, and the values
presented in \cite{dickeyb} for clusters at lower declinations. In
order to sample the observed distribution of cluster redshifts and
column densities, and the expected distribution of cluster
temperatures, we derived conversion factors over the following ranges;
({\it i}) $0.06<z<0.86$ (in increments of $\delta z=0.05$), ({\it ii})
0$<$nH$<$20$\times10^{20}$ cm$^{-2}$ (in increments of
$1\times10^{20}$ cm$^{-2}$), and ({\it iii}) 1$<$T$<$12 keV (in
increments of 1 keV). (When a cluster redshift or column density was
not exactly matched by one of the matrix entries, linear interpolation
was used.) As expected, the count rate to flux conversion varied most
rapidly along the column density axis of this matrix, however changing
the redshift also had a measurable effect (by a factor of $\simeq 2$
over the range 0.08$<z<$0.8). Estimates of the bolometric\footnote{An
energy range of 0.01-50 keV was used to calculate the (pseudo)
bolometric corrections, which were found to be in excellent agreement
with those presented in Figure 2 of \cite{Borgani}.} and k-corrections
were also derived, as a function of temperature, using XSPEC.

The luminosity derivation included an iteration to obtain an estimate
of the X-ray temperature for each cluster, using the
luminosity--temperature (L-T) relation presented in
\cite{ArnaudMb}. From a starting point of T=6 keV an initial [0.5-2.0
keV] luminosity was derived. This luminosity was then converted into a
pseudo bolometric luminosity, so that a temperature estimate (to the
nearest integer in keV) could be derived.  The new temperature was
used to select a second count rate to flux conversion from the matrix
and the process was repeated until convergence was reached. The
temperature used in the final luminosity calculation is listed in
column 14 of Table \ref{clusters}.

In the past, the luminosity-temperature (L-T) relation was not so well
known and other groups have adopted a single temperature, usually 6
keV, for their luminosity calculations. Using the \cite{ArnaudMb} L-T
relation, 6 keV corresponds to a cluster of $L_{44}\simeq 6$. Most of
the clusters in Table \ref{clusters} are significantly fainter than
this, meaning that the use of a canonical temperature will yield
inaccurate results, especially for the lowest luminosity clusters.
This is illustrated by the faintest (and hence, coolest) cluster in
our sample (RX J1524.6) which has a luminosity of $L_{44}$=0.065 when
a temperature of T=1 keV is assumed and a luminosity of $L_{44}$=0.072
when a temperature of T=6 keV is assumed (an 11\% effect).  By
contrast the effect is smaller (5\%) for the hottest cluster in our
sample; RX J0152.7 has a luminosity of $L_{44}=8.26$ when a
temperature of T=9 keV is assumed and a luminosity of $L_{44}$=8.65
when a temperature of T=6 keV is assumed. It is worth mentioning that
the L-T relation we use (\cite{ArnaudM}) was constructed from clusters
known not to contain cooling flows. Another recent work (\cite{Allen})
combines both non-cooling flow and cooling flow clusters and fits a
flatter slope to the L-T relation (2.4 compared to
2.9). Unfortunately, the poor photon statistics of the Bright SHARC
cluster sample do not allow us to test for the presence of cooling
flows and so our choice of L-T relation will be inappropriate in some
cases.

Finally, we note that the conversion between cluster count rate and
cluster luminosity is a function of the adopted values of Hubble's
Constant and the deceleration parameter and that we have used H$_0$=50
km s$^{-1}$ Mpc$^{-1}$ and q$_0$=0.5 throughout.

\section{Discussion}
\label{results}

In a companion paper (\cite{Nichol99}, N99 hereafter) we use the
Bright SHARC sample to examine evolution in the X-ray cluster
luminosity function (XCLF). Future papers will go on to use these
evolution results to constrain the density parameter $\Omega_m$. It is
appropriate, therefore, to discuss here some of the observational
issues relevant to $\Omega_m$ analyses.  These issues include
systematic biases in the derived luminosities (\S\ref{luminbias}) of
the Bright SHARC clusters and any possible contamination
(\S\ref{contam}), or incompleteness (\S\ref{complete}) in the Bright
SHARC catalog. We also discuss the possible discovery of three
``fossil groups'' (\S\ref{fossil}) and our overlap with the 160
deg$^2$ survey of V98 (\S\ref{V98}).

\subsection{Luminosity Bias in the Bright SHARC Cluster Sample}
\label{luminbias}

A systematic bias in our luminosities would result in an over (or
under) estimate of the number density of high luminosity systems. To
investigate whether such a systematic bias exists, we have compared
the luminosities quoted in column 13 of Table \ref{clusters} with
published values for the six clusters we have in common with the EMSS
(\cite{gioia90}): RX J1024.3 (MS1020.7 or A981), RX J1204.0
(MS1201.5), RX J1211.2 (MS1208.7), RX J1222.1 (MS 1219.9), RX J1311.2
(MS1308.8), RX J2258.1 (MS2255.7 or Z2255.5). We have chosen to
compare our luminosities for these clusters with those presented in
\cite{Nichol97bc}, N97 hereafter), rather than those presented in
\cite{henry92b}, for two reasons. First the luminosities quoted 
in N97 are in the ROSAT bandpass [0.5-2.0 keV] rather than the
Einstein bandpass [0.3-3.5 keV]. Second, the N97 study used the SHARC
pipeline to produce count rate maps.  A comparison of the two sets of
luminosities will, therefore, show if the methodology outlined in
sections \S\ref{aperture} and \S\ref{convert} is robust (since a
different methodology\footnote{Differences between Bright SHARC and
N97 include the use of the IRAF PROS package to set background
apertures and the use of a constant temperature, T=6 keV, for
k-corrections and count rate to flux conversions.} was used in
N97). We find our derived luminosities to differ by 0\% for MS1020.7,
1\% for MS1201.5, 5\% for MS1208.7, 26\% for MS1219.9, 2\% for
MS1303.8 and 10\% for MS1219.9. These differences are all smaller than
the 1 sigma errors on the EMSS count rates quoted in N97. We note also
that the luminosity we derive for RX J0152.7 is within 1\% of the
value derived by the WARPS collaboration in \cite{Ebeling99b}.

We have also compared the fluxes quoted in column 12 of Table
\ref{clusters} with published values for the 11 clusters we have in
common with the 160 deg$^2$ survey (V98, see \S\ref{complete}). In
Table \ref{vik_comp}, column 5, we present the ratio of Bright SHARC
to V98 fluxes. We find the Bright SHARC values to be systematically
higher than those measured by V98, with an average flux ratio of
1.18. To understand this discrepancy, we have recalculated the Bright
SHARC fluxes using the core radii and redshifts presented in
V98. Except for RX J1641.2, the V98 core radii are all smaller than
$r_c$=250kpc and, by using their values, we bring the average flux
ratio down to 1.01 (column 6).

We conclude that the methodology of sections \S\ref{aperture} \&
\S\ref{convert} is robust, although it has the disadvantage of over estimating
the cluster flux if $r_c<$250kpc.  Planned XMM observations of several
Bright SHARC clusters will provide higher angular resolution and
signal-to-noise images together with accurate estimates of the
electron temperature. These observations will provide an important
test of the methods described in sections \S\ref{aperture} \&
\S\ref{convert} since they will allow us to ({\it i}) more accurately
excise contaminating sources in the cluster aperture, ({\it ii}) use
fitted, rather than canonical, values for $\beta$, $r_c$ and the
ellipticity, ({\it iii}) be less sensitive to errors in the background
calculation and ({\it iv}) improve our spectral dependent count rate
to flux conversions.

\subsection{Contamination of the Bright SHARC Cluster Sample}
\label{contam}

The thorough, multi-object, spectroscopic follow-up of the Bright
SHARC extended source list means that it is highly unlikely that any
of the entries in Table\ref{clusters} are mis-identified contaminants.
However, we stress that there are two clusters in that table which
should {\it not} be used for studies of the cluster XCLF because their
detections are not truly serendipitous: RX J1024.3 and RX J1541.1 were
found in cluster pointings and lie at redshift separations from the
pointing target of $\delta z<0.002$, or $cz<$600 km s$^{-1}$. These
clusters are probably associated with the pointing target via the
cluster correlation function (\cite{romer94}; \cite{nichol94mn}). In
addition, we feel that RX J1222.1 (MS1219.9) warrants further study:
This object is very compact, has a large count rate uncertainty
(\S\ref{aperture}) and \cite{GioiaLb} note that its central galaxy has
emission lines. It is possible, therefore, that the luminosity quoted
in Table \ref{clusters} is an overestimate due to AGN contamination.
(Although, it should be noted that the presence of emission lines in
the central galaxy could be attributed to cooling flow nebulosity or
star formation, \cite{Crawford}.) We note that the three clusters
highlighted here (RX J1024.3, RX J1222.1 \& RX J1541.1) have redshifts
in the range $0.20<z<0.25$ and so were not used in the N99 analysis
(which concentrated only on those clusters at $z>$0.3).

\subsection{Incompleteness of the Bright SHARC Cluster Sample}
\label{complete}
There are three possible ways in which the Bright SHARC cluster sample
might be incomplete. First there are those clusters that did not meet
our selection criteria. Second, there is a possibility that some
clusters were misidentified as contaminants.  Third, there are the
three extended sources which have yet to be identified.

We are using simulations to understand how the adopted selection
criteria (\S\ref{extent}) effects the completeness of the Bright SHARC
cluster sample.  We are in the process of carrying out a very thorough
investigation of our selection function by adding many thousands of
fake clusters (one at a time) to the pointings in our survey and then
determining the fraction of these fake clusters that would have been
selected as Bright SHARC cluster candidates. These simulations will
provide us with the efficiency of cluster detection as a function of
cluster parameters (\eg redshift, luminosity, ellipticity, core radius
\etc) and operational parameters (\eg exposure time, off-axis angle,
Hydrogen column density, central target \etc). The results of these
simulations will be presented elsewhere (\cite{adami}), but our
preliminary findings are described in N99.

Let us now address possible cases where clusters might be
misidentified as contaminants. We discuss first the two objects listed
in Table \ref{XrayBrightSHARC} as blends of a cluster with another
source, RX J0318.2 \& RX J2314.7. The former, RX J0318.2, is a blend
of a cluster with a QSO. (The cluster has the same redshift as the
neighboring cluster RX J0318.8, $z$=0.37).  The surface brightness
contours of RX J0318.2 are clearly dumb-bell shaped and so it has been
possible to remove the QSO contribution from the total count
rate. This object was also discovered as part of the Southern SHARC
and \cite{burkethesisb} has determined the total count rate and
luminosity of this cluster to be $cr_T$=0.01362 count s$^{-1}$ and
$L_{44}=1.11$ respectively.  Therefore, this cluster would not have
made it into the Bright SHARC sample had it not been blended with the
QSO and its exclusion for Table \ref{clusters} is justified.  By
contrast, the boundary between the cluster and M-star emission for RX
J2314.7 is blurred. Hence it is not possible to excise the M-star flux
to see if the cluster alone has a high enough count rate (and extent)
to qualify as a Bright SHARC candidate. If the M-star makes only a
minimal contribution, less that 20\%, to the total flux, then the
cluster should have been included in Table \ref{clusters}: Assuming
that all the RX J2314.7 flux comes from the cluster, the cluster would
have a luminosity of $L_{44}$=1.31.

As stated above, three of the 94 Bright SHARC extended sources remain
unidentified.  If all three were high redshift, high luminosity
clusters, then there would be important implications for cluster
evolution. In N99, we predict that the Bright SHARC survey should
include 4.9 clusters with luminosities $L_{44}\geq5$ in the redshift
range $0.3<z<0.7$ (based on a simple extrapolation of the
\cite{DeGrandia} local XCLF). Since only 1 such cluster has been
confirmed to exist in the Bright SHARC (RX J1120.1), we conclude in
N99 that there may be evidence for evolution at luminosities brighter
than $L_{44}=5$. This evidence would effectively disappear if another
3 Bright SHARC clusters were added in this luminosity range. We
stress, however, that it is very unlikely all these objects are
clusters with luminosities brighter than $L_{44}=5$; the CCD images of
RX J0340.1 \& RX J1705.6 are not consistent with the presence of
distant clusters and RX J1838.8 is in a crowded star field (and so is
most likely associated with a stellar X-ray source). We conservatively
estimate that one these objects may be a cluster, given that the ratio
of clusters to non-clusters among the other 91 identified sources is
roughly 1:3.  We have calculated that this cluster would have to
reside at $z>0.62$, $z>0.57$ or $z>0.51$, for RX J0340.1, RX J1705.6
and RX J1838.8 respectively, to have a luminosity greater than
$L_{44}=5$.

We also highlight candidate RX J1210.4. This object contains a QSO and
has a compact X-ray surface brightness profile. Even though most of
the flux from this source is probably coming the QSO, this object
merits further study since a CCD image highlights a clustering of
faint galaxies around the bright central object. The redshift of this
source ($z$=0.615) and its high count rate ($cr_W=0.1430$) mean that
if more than 18\% of the count rate from this source was coming from
an associated cluster, then this cluster would have a luminosity
greater than $L_{44}$=5.

For the various reasons outlined above, we have decided to continue
the follow-up of the Bright SHARC in a variety of ways. As a first
priority, we plan to identify the three remaining unidentified Bright
SHARC extended sources (RX J0340.1, RX J1705.6 and RX J1838.8). We
also plan to obtain identifications for at least one portion of the
seven ``id-pending'' blends listed in Table \ref{XrayBrightSHARC} and
to continue our campaign to obtain velocity dispersions for the Bright
SHARC clusters. Moreover, we hope to obtain higher resolution X-ray
images of complex sources such as RX J1210.4, RX J1222.1 and RX
J2314.7, to help determine the contamination level.

\subsection{Fossil Groups and Dark Clusters in the Bright SHARC Survey}
\label{fossil}

We present evidence for the discovery of three new ``fossil groups''
(\cite{ponman94}) or X-ray Over-Luminous Elliptical Galaxies (OLEGs,
\cite{Viklinin99}).  These objects are predicted to occur when a
galaxy group relaxes to form a single elliptical galaxy. They are
interesting because they provide invaluable insight into the processes
of elliptical galaxy evolution, metal enrichment in the intra cluster
medium, and the dynamics of extended dark halos (\cite{mulchaey}).
Their observational signatures would be an isolated cD or giant
elliptical galaxy surrounded by a cool (T$\simeq$1 keV), extended,
X-ray halo.  Two galaxies detected in the Bright SHARC survey appear
to share these properties; RX J1730.6 (NGC6414, z=0.05) and RX J0327.9
(UGC2748, z=0.03).  Applying the same method used to obtain total
cluster count rates (\S\ref{aperture}), we have measured their
luminosities to be $L_{44}$=0.158 and $L_{44}$=0.056
respectively\footnote{Assuming an absorbed Raymond Smith spectrum with
an electron temperature of T=1 keV.}.  In addition to these two
galaxies, one of the Bright SHARC clusters, RX J0321.9 (A3120,
$z$=0.0696, $L_{44}=0.43$), also appears to display ``fossil group''
characteristics. We highlight these objects here since they are ideal
targets for follow-up studies at X--ray and optical wavelengths.  We
have estimated the ``fossil group'' space density to be $\sim2\times
10^{-6}{\rm Mpc^{-3}}$ under the assumption that the Bright SHARC is
100\% efficient in detecting extended sources in the redshift range
$0.02<z<0.08$ and at luminosities of $L_{44}>$0.1.

In addition to estimating the space density of ``fossil groups'', we
can comment on the space density of ``dark clusters'' or ``failed
clusters''. These objects are theorized to have cluster-like masses,
and to radiate in the X--rays, but to have an under luminous galactic
component (\cite{tucker}; \cite{Hattori}).  We have successfully
identified 91 of the 94 Bright SHARC extended sources and have found
no evidence for ``dark clusters''. To avoid detection in the Bright
SHARC, these objects either must have a lower space density than rich
clusters and ``fossil groups'', or they must be intrinsically faint
and evolve rapidly (to avoid detection at low redshift). In either
case, ``dark clusters'' are unlikely to be a significant contribution
to the mass of the universe.

\subsection{Comparison with the 160 deg$^2$ Survey}
\label{V98}

\begin{figure}
\centerline{\psfig{figure=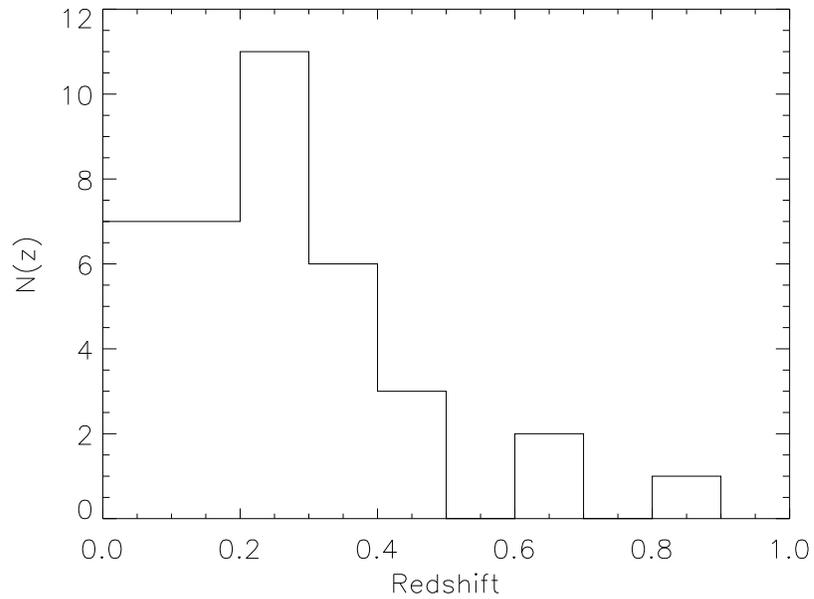,height=3.5in,angle=90}}
\caption{Redshift distribution of the thirty-seven clusters in
the Bright SHARC.}
\label{zhisto}
\end{figure}

As pointed out by N99, it may be possible to combine the Bright SHARC
with the 160 deg$^2$ survey (V98), to maximize the area available for
high redshift cluster searches. The motivation for this is
demonstrated by Figure \ref{zhisto}, which shows several gaps in the
redshift coverage of our survey. Even though we are able to find high
luminosity clusters out to at least $z=0.83$, we find none at
$z\simeq0.5$ or $z\simeq0.7$.  The only way to guarantee more
$L_{44}>3$ cluster detections would be to search over a wider
area. The combination of the two surveys would yield a search area of
$\simeq$ 260 deg$^2$, since only 44\% (or $\simeq$ 78 deg$^2$) of the
Bright SHARC Survey overlaps with the 160 deg$^2$ survey. (There are
201 pointings in common between the 160 deg$^2$ and Bright SHARC
Surveys; Alexey Vikhlinin, private communication).

There are 13 sources in common between the Bright SHARC and the 160
deg$^2$ surveys.  Of these 13, five clusters have not been followed up
spectroscopically by either survey but rely on literature redshifts
(RX J1010.2\footnote{RX J1010.2 was not included in Table
\ref{clusters} because its redshift ($z$=0.045) is too low, \ie
$z$$<$0.07.}, RX J1204.0, RX J1211.2, RX J1311.2, RX J2258.1). An
additional three clusters have both Bright SHARC and V98 redshifts (RX
J0849.1, RX J1406.9, RX J1701.3); with the two redshifts being in
agreement in all cases. We have also been able to provide
spectroscopic information for five 160 deg$^2$ sources which
previously relied on photometric redshifts; RX J0237.9 (V28), RX
J0947.8 (V75), RX J1418.5 (V159), RX J1524.6 (V170), RX J1641.2
(V183). We have identified RX J0947.8 as a blend, the main component
of which is a QSO\footnote{Subsequent observations by Vikhlinin \et
have shown that this QSO most likely resides on the outskirts of a
cluster at the same redshift (Alexey Vikhlinin, private
communication).}  at $z$=0.63 (\cite{burkethesis}).  We confirm that
the other four sources are clusters and we find that the photometric
redshifts listed in V98 are good estimates of the true redshift, with
the largest error being $\delta z=0.065$ for RX J1641.2.  This cluster
has been shown to be at $z=0.195$, giving it a luminosity of
$L_{44}=1.355$.  It is not, therefore, a high redshift, high
luminosity, cluster, as previously suggested by \cite{Viklinin2bc}),
based on the upper limit of the estimated redshift
($z_{est}=0.26^{+0.04}_{-0.07}$).

In addition to the 13 sources described above, 77 other V98 clusters
were detected in the 201 pointings common to the two surveys. Most of
these clusters are too faint to have been included in the Bright SHARC
sample, only 9 have wavelet count rates greater than the Bright SHARC
threshold ($cr_W=0.01163$).  Of these 9, seven were not included in
the Bright SHARC because they did not meet our filling factor
criterion ($f<1.3$), one was detected at an offaxis distance less than
our threshold of $2'.5$ and one did not meet our extent criterion.
Conversely, two clusters (RX J0256.5 and RX J1311.8) in Table
\ref{clusters} are not listed in V98, despite falling in common
pointings, because they lie beyond the V98 offaxis limit of
$17'.5$. These examples demonstrate how differing survey selection
criteria produce differing cluster samples and that detailed
simulations are required to determine a survey's selection function.

There are eight confirmed $L_{44}>3$ clusters in the Bright SHARC; RX
J0152.7, RX J0256.5, RX J0318.5, RX J0426.1, RX J1241.5, RX J1120.1,
RX J1334.3, RX J1701.3. The presence of so many $L_{44}>3$ clusters in
the Bright SHARC has allowed us to show that the XCLF is non-evolving
up to $L_{44}\simeq5$ (N99). It is important to note that, even after
the combination of Bright SHARC and 160 deg$^2$ surveys, the areal
coverage available for high redshift cluster searches will still be
only about one third that of the EMSS at the bright end
(\cite{henry92}). This means that we will probably have to wait until
larger area surveys are made available, e.g., from the XMM satellite 
(\cite{romer2000}), to make definitive statements about XCLF
evolution at $L_{44}>5$.\\

{\bf Appendices A through E have been ommitted from the astro-ph
submission.\footnote{The appendices are accessible from
http://www.journals.uchicago.edu/ApJ/journal/issues/ApJS/v126n2/40418/40418.html}}

\vspace{0.25in}
\centerline{\bf Acknowledgements}
\vspace{0.25in}

We acknowledge financial support from NASA grants NAG5-2432 (AR, RP,
CA \& MU), NAG5-6548 (RN), NAG5-3202 (BH), GO-06838.01-95A (BH). And
also from a NASA Space Consortium Grant through Aerospace Illinois
(AM, KC, BH), the CMU undergraduate research initiative (AJM), the NSF
Center for Astrophysical Research in Antartica (BH), NSF grant
AST-9256606 (BH) and PPARC (DB).  This research has made use of:
%HEASARC
({\it i}) Data obtained through the High Energy Astrophysics Science 
Archive Research Center Online Service, provided by the NASA-Goddard 
Space Flight Center.
%DSS
({\it ii}) The Digitized Sky Survey which was produced at the Space
Telescope Science Institute under US Government grant NAG W-2166.
%NED  
({\it iii}) The NASA/IPAC Extragalactic 
Database (NED) which is operated by the Jet Propulsion Laboratory, 
Caltech, under contract with the National Aeronautics and Space Administration.
%APM at AAO
({\it iv}) The ``APM Catalogues at AAO'' web server, author Micheal Drinkwater.
%APS
({\it v}) The APS Catalog of POSS I and the APS Image Database, which
are supported by the National Science Foundation, the National
Aeronautics and Space Administration, and the University of Minnesota.
We offer special thanks to Jim DeVeny and the support staff at the
ARC, CFHT, CTIO, ESO \& KPNO telescopes, and also to Alain Blanchard,
Francisico Castander, Ian Del Antonio, Paul Lynam, Eric Monier,
Francis Falbo, Tim Kimball, Marc Postman, Patricia Purdue, Connie
Rockosi, Rachid Sadat, Steve Snowden, Jeffrey Tran, Dave Turnshek,
Pedro Viana, Alexey Vikhlinin and an anonymous referee.

%\addtocounter{page}{-1}
%Latex unfortunately leaves this page blank.

%\documentstyle [psfig,aaspp,supertab,side,11pt]{article}
%\begin{document}
\small{
\begin{sidetable}
%\addtocounter{table}{-1}
\caption{extended sources in the Bright SHARC survey}
\begin{center}
\begin{tabular}{llrrlllcr} \hline\hline
\multicolumn{1}{c}{Source} & \multicolumn{2}{c}{RA (J2000) Dec} & \multicolumn{1}{c}{$cr_W$} & \multicolumn{1}{c}{Pointing} & \multicolumn{1}{c}{ID} & \multicolumn{1}{c}{ID code} & \multicolumn{1}{c}{Notes} & \multicolumn{1}{c}{Ref}\\ 
\multicolumn{1}{c}{(1)}&\multicolumn{1}{c}{(2)}&\multicolumn{1}{c}{(3)}&\multicolumn{1}{c}{(4)}&\multicolumn{1}{c}{(5)}&\multicolumn{1}{c}{(6)}&\multicolumn{1}{c}{(7)}&\multicolumn{1}{c}{(8)}&\multicolumn{1}{c}{(9)} \\	\hline
RX J0031.0--3547 & 00 31 03.0 & --35 47 21.8 & 1.26 & wp800387n00 & Blend   &  X S N C  & AGN, $z$=0.25 + ?& \\ 
RX J0031.9--3556 & 00 31 59.5 & --35 56 12.1 & 1.19 & wp800387n00 & Blend   &  X S C  & id pending& \\ 
RX J0058.0--2721 & 00 58 00.5 & --27 21 29.3 & 1.18 & rp701223n00 & Blend   &  S N C  & Mstar + ?& \\ 
RX J0117.6--2238 & 01 17 36.5 & --22 38 13.5 & 1.25 & rp100376n00 & Cluster &  S N C  & A2894, $z$=0.207& \\ 
RX J0124.8+0932 & 01 24 48.2 & +09 32 30.8 & 1.92 & rp700976    & Galaxy    &  N      & NGC 524, $cz$=2421 km/s &1\\  
RX J0152.7--1357 & 01 52 42.0 & --13 57 52.9 & 1.71 & rp600005n00 & Cluster &  S O C  & $z$=0.83 &2\\   
RX J0209.4--1008 & 02 09 24.2 & --10 08 04.2 & 1.34 & rp800114n00 & Galaxy  &  N  & NGC0835, $z$=0.0135 &3\\  
RX J0209.9--1003 & 02 09 58.1 & --10 03 19.2 & 1.47 & rp800114n00 & Blend   &  X S N C  & QSO: MS0207.4, $z$=1.97 + ? &4\\  
RX J0217.4--1800 & 02 17 26.1 & --18 00 05.7 & 1.38 & rp900352n00 & Blend   &  S C  &  AGN, $z$=0.345 + ?& \\ 
RX J0221.1+1958 & 02 21  08.4 & +19 58 25.9 & 1.54 & wp900147    & Cluster   &  S C  &  $z$=0.45& \\ 
RX J0223.4--0852 & 02 23 28.1 & --08 52 14.3 & 1.20 & rp800016n00 & Cluster &  S O N C  & $z$=0.163 &5\\   
RX J0237.9--5224 & 02 37 59.1 & --52 24 45.7 & 2.32 & rp300201n00 & Cluster &  S O N C  & A3038, $z$=0.133 &6\\   
RX J0250.0+1908 & 02 50  02.9 & +19 08 29.4 & 1.42 & rp700920    & Cluster   &  S C  &  $z$=0.12& \\ 
RX J0256.5+0006 & 02 56 32.9 & +00 06 11.6 & 3.61 & rp701403n00 & Cluster   &  S C  &  $z$=0.36& \\ 
RX J0318.2--0301 & 03 18 17.3 & --03 01 21.1 & 1.33 & wp800555n00 & Blend   &  X S C  & {\scriptsize Cluster, $z$=0.37 + AGN, $z$=0.233}& \\ 
RX J0318.5--0302 & 03 18 33.3 & --03 02 46.7 & 2.76 & wp800555n00 & Cluster &  S C  &   $z$=0.37& \\ 
RX J0321.9--5119 & 03 21 57.0 & --51 19 33.1 & 6.18 & wp800371n00 & Cluster &  N  & A3120, $z$=0.0696 &7\\  
RX J0324.6--5103 & 03 24 37.9 & --51 03 52.1 & 1.32 & wp800371n00 & Blend   &  X D  & Star:HD21360 + ?& \\ 
RX J0327.9+0233 & 03 27 54.3 & +02 33 43.2 & 3.23 & rp700099m01 & Galaxy    &  N  & UGC2748, $z$=0.0302 &8\\  
RX J0337.5--2518 & 03 37 34.2 & --25 18 01.5 & 2.13 & wp300079    & Blend   &  X C  & id pending& \\ 
RX J0340.1--4458 & 03 40 09.0 & --44 58 48.2 & 1.24 & rp900495n00 & Pending  &  C  & & \\ 
RX J0359.1--5300 & 03 59 11.9 & --53 00 56.2 & 1.51 & rp800308    & Blend   &  X D  & 2 stars& \\ 
RX J0414.0--1224 & 04 14 05.7 & --12 24 24.9 & 2.50 & rp900242n00 & Blend   &  S N  & AGN, $z$=0.572 + ? &9\\   
RX J0415.7--5535 & 04 15 45.4 & --55 35 31.0 & 1.20 & wp600623n00 & Galaxy  &  N  & NGC1549, $cz$=1197 km/s &10\\  
RX J0416.1--5546 & 04 16 10.3 & --55 46 46.3 & 3.65 & wp600623n00 & Galaxy  &  N  & NGC1553, $cz$=1080 km/s &11\\ 
RX J0420.9+1444 & 04 20 58.7 & +14 44 07.7 & 2.83 & wp200441    & Blend   &  S C  & AGN + ?&\\ 
RX J0421.2+1340 & 04 21 16.8 & +13 40 14.8 & 1.32 & rp200776n00 & Blend   &  X S C  & star + ?& \\ 
RX J0426.1+1655 & 04 26 07.3 & +16 55 12.1 & 1.79 & rp201369n00 & Cluster &  S C  &  $z$=0.38& \\ 
RX J0454.3--0239 & 04 54 19.6 & --02 39 50.0 & 1.59 & rp800229n00 & Cluster   &  S C  & $z$=0.26& \\ 
RX J0514.2--4826 & 05 14 16.7 & --48 26 53.4 & 1.58 & wp800368n00 & Blend &  X S C  & AGN, $z$=0.230 + ?& \\ 
RX J0609.1--4854 & 06 09 06.5 & --48 54 50.4 & 2.15 & rp300111    & Blend &  D  & star + ?& \\ 
RX J0849.1+3731 & 08 49 08.9 & +37 31 47.9 & 1.29 & rp700546n00 & Cluster &  S N C  & A708, $z$=0.23 &\\  
RX J0853.6+1349 & 08 53 41.1 & +13 49 29.5 & 1.17 & rp700887n00 & Blend   &  X D N  & {\scriptsize Star + Galaxy:MS0850.8, $z$=0.194} &12\\ 
RX J0945.6--1434 & 09 45 40.4 & --14 34  5.0 & 2.30 & wp701458n00 & Blend &  X D S  & AGN, $z\simeq$1.2 + star& \\ 
RX J0947.8+0741 & 09 47 50.5 & +07 41 43.0 & 1.51 & wp701587n00 & Blend   &  S C  &  QSO, $z\simeq$0.63 + ?& \\ 
RX J1010.2+5430 & 10 10 12.9 & +54 30 09.6 & 1.41 & wp900213    & Group   &  N  & V84 $z\simeq0.045$ &13\\  
RX J1020.0+3915 & 10 20 02.4 & +39 15 50.8 & 1.18 & wp900528n00 & Blend   &  X C  & id pending& \\ 
RX J1024.3+6805 & 10 24 20.1 & +68 05 05.1 & 2.60 & wp800641n00 & Cluster &  N  & A981, $z$=0.201 &14\\  
RX J1031.3--1433 & 10 31 23.3 & --14 33 40.6 & 1.18 & rp700461n00 & Blend &  D S  &  Star + ?& \\ 
RX J1113.8+4017 & 11 13 48.5 & +40 17 18.3 & 1.44 & rp700855n00 & Cluster &  N  &  A1203, $z$=0.0795 &15\\ 
RX J1120.1+4318 & 11 20 07.5 & +43 18 04.9 & 2.11 & rp900383n00 & Cluster &  S C  & $z$=0.60& \\ 
RX J1142.2+1026 & 11 42 16.7 & +10 26 46.9 & 1.27 & wp600420    & Cluster &  N  &  A1356, $z$=0.0698 &16\\  
RX J1143.7+5520 & 11 43 46.5 & +55 20 13.6 & 1.98 & rp600236n00 & Blend   &  X D C  & id pending& \\ 
\end{tabular}
\end{center}
\end{sidetable}
\begin{sidetable}
\addtocounter{table}{-1}
\begin{center}
\caption{(continued)}
\begin{tabular}{llrrlllcr} \hline\hline
\multicolumn{1}{c}{Source} & \multicolumn{2}{c}{RA (J2000) Dec} & \multicolumn{1}{c}{$cr_W$} & \multicolumn{1}{c}{Pointing} & \multicolumn{1}{c}{ID} & \multicolumn{1}{c}{ID code} & \multicolumn{1}{c}{Notes} & \multicolumn{1}{c}{Ref}\\ 
\multicolumn{1}{c}{(1)}&\multicolumn{1}{c}{(2)}&\multicolumn{1}{c}{(3)}&\multicolumn{1}{c}{(4)}&\multicolumn{1}{c}{(5)}&\multicolumn{1}{c}{(6)}&\multicolumn{1}{c}{(7)}&\multicolumn{1}{c}{(8)}&\multicolumn{1}{c}{(9)} \\	\hline
RX J1204.0+2807 & 12 04 03.6 & +28 07 03.6 & 4.49 & wp700232    & Cluster &  N  & MS1201.5, $z$=0.167 &17\\   
RX J1204.1+2020 & 12 04 09.7 & +20 20 40.5 & 2.04 & rp800039    & Galaxy  &  N  & NGC4066, $z$=0.024 &18\\   
RX J1210.4+3929 & 12 10 25.9 & +39 29 07.6 & 14.30 & wp700277    & Blend   &  N C  &  {\tiny QSO:MS1207.9, $z$=0.615 + ?} &19\\   
RX J1211.1+3907 & 12 11 09.5 & +39 07 44.4 & 2.14 & rp600625n00 & Blend   &  X D  & Star + ?& \\
RX J1211.2+3911 & 12 11 14.5 & +39 11 41.1 & 1.52 & wp700277    & Cluster &  N  &   MS1208.7, $z$=0.34 &20\\   
RX J1220.3+7522 & 12 20 18.0 & +75 22 10.2 & 4.65 & rp700434    & Galaxy  &  N  &   NGC4291, $z$=0.059 &21\\   
RX J1222.1+7526 & 12 22 06.9 & +75 26 16.8 & 1.17 & rp700434    & Cluster &  N  &  MS1219.9, $z$=0.24 &22\\  
RX J1222.5+2550 & 12 22 30.8 & +25 50 26.7 & 36.93 & wp200307    & Blend   &  X D  & 2 Stars& \\ 
RX J1227.4+0849 & 12 27 27.6 & +08 49 53.1 & 5.87 & wp600587n00 & Cluster &  N    & A1541, $z$=0.0895 &23\\   
RX J1232.8+2605 & 12 32 48.3 & +26 05 39.0 & 1.82 & rp600162    & Cluster &  S C  & $z$=0.22&\\ \hline
RX J1241.5+3250 & 12 41 33.1 & +32 50 22.9 & 2.38 & rp600129a00 & Cluster &  S C  & $z$=0.39& \\ 
RX J1244.1+1134 & 12 44 08.2 & +11 34 16.8 & 1.16 & rp600017    & Blend   &  X C  & id pending& \\ 
RX J1250.4+2530 & 12 50 26.1 & +25 30 17.6 & 1.71 & wp900212    & Galaxy  &  N    & NGC4725, $z$=0.00402 &24\\   
RX J1259.7--3236 & 12 59 45.4 & --32 36 59.9 & 1.19 & rp800384n00 & Cluster  &  S C  & $z$=0.076 & \\ 
RX J1308.5+5342 & 13 08 32.6 & +53 42 19.3 & 1.25 & wp300394n00 & Cluster &  S C  & $z$=0.33& \\  
RX J1311.2+3228 & 13 11 12.3 & +32 28 53.2 & 2.53 & wp700216    & Cluster &  N C  & MS1308.8, $z$=0.245 &25\\   
RX J1311.8+3227 & 13 11 49.8 & +32 27 40.4 & 1.47 & wp700216    & Cluster &  S C  & $z$=0.44& \\ 
RX J1313.6--3250 & 13 13 41.0 & --32 50 45.9 & 1.28 & wp300219  & Blend   &  X D  & id pending& \\ 
RX J1334.3+5030 & 13 34 20.0 & +50 30 54.2 & 1.36 & rp800047    & Cluster &  S C  & $z$=0.62& \\ 
RX J1343.7+5538 & 13 43 45.2 & +55 38 20.3 & 1.80 & rp700922n00 & Cluster &  N  & A1783, $z$=0.0766 &26\\   
RX J1349.2--0712 & 13 49 12.3 & --07 12 41.2 & 1.51 & rp800637n00 & Gal. pair    &  N  & part of HCG67, z=0.02406 &27\\  
RX J1406.9+2834 & 14 06 55.1 & +28 34 15.7 & 1.30 & rp700061    & Cluster &  N S C & V154, $z$=0.117 &\\  
RX J1412.4+4355 & 14 12 29.8 & +43 55 31.2 & 2.32 & wp700248    & Blend   &  X S N C  & AGN, $z$=0.095 + ? &28\\  
RX J1416.4+2315 & 14 16 26.6 & +23 15 32.8 & 4.80 & rp800401a01 & Cluster &  S C  & $z$=0.138& \\ 
RX J1417.9+5417 & 14 17 57.5 & +54 17 51.3 & 1.26 & wp150046    & Blend   &  O C  & AGN + Mstar &29\\   
RX J1418.5+2510 & 14 18 31.4 & +25 10 45.8 & 3.78 & wp150071    & Cluster &  S C  & V159, $z$=0.29& \\ 
RX J1508.4+5537 & 15 08 24.6 & +55 37 05.3 & 1.16 & rp600119n00 & Blend   &  D S C& Star + ?& \\ 
RX J1517.1+3140 & 15 17 08.4 & +31 40 58.4 & 1.29 & rp201018    & Blend   &  X D S C  & Star + ?& \\ 
RX J1524.6+0957 & 15 24 39.6 & +09 57 44.8 & 1.64 & rp701001n00 & Cluster &  S O C& V170, $z$=0.078 &30\\   
RX J1525.3+4201 & 15 25 23.3 & +42 01 00.0 & 1.26 & rp701405n00 & Blend   &  N C  & QSO, $z$=1.189 + ? &31\\  
RX J1541.1+6626 & 15 41 10.3 & +66 26 25.0 & 1.51 & rp800511n00 & Cluster &  S C  & $z$=0.245& \\ 
RX J1543.7+6627 & 15 43 42.7 & +66 27 42.3 & 1.25 & rp800511n00 & Blend   &  S C  & QSO, $z$=1.4562 + ?& \\ 
RX J1641.2+8233 & 16 41 13.9 & +82 33 01.7 & 3.55 & rp700098    & Cluster &  S N C  & V183, $z$=0.195 &32\\   
\end{tabular}
\end{center}
\end{sidetable}
\begin{sidetable}
\addtocounter{table}{-1}
\begin{center}
\caption{(continued)}
\begin{tabular}{llrrlllcr} \hline\hline
\multicolumn{1}{c}{Source} & \multicolumn{2}{c}{RA (J2000) Dec} & \multicolumn{1}{c}{$cr_W$} & \multicolumn{1}{c}{Pointing} & \multicolumn{1}{c}{ID} & \multicolumn{1}{c}{ID code} & \multicolumn{1}{c}{Notes} & \multicolumn{1}{c}{Ref}\\ 
\multicolumn{1}{c}{(1)}&\multicolumn{1}{c}{(2)}&\multicolumn{1}{c}{(3)}&\multicolumn{1}{c}{(4)}&\multicolumn{1}{c}{(5)}&\multicolumn{1}{c}{(6)}&\multicolumn{1}{c}{(7)}&\multicolumn{1}{c}{(8)}&\multicolumn{1}{c}{(9)} \\	\hline
RX J1701.3+6414 & 17 01 22.5 & +64 14 08.3 & 1.98 & wp701457n00 & Cluster &  N S  C  & V190, $z$=0.453 &33\\   
RX J1705.6+6024 & 17 05 37.5 & +60 24 11.0 & 1.46 & rp701439n00 & Pending   &  C  && \\ 
RX J1726.2+0410 & 17 26 14.4 & +04 10 23.8 & 1.98 & rp200522n00 & Blend   &  X D  & Star + ?& \\ 
RX J1730.6+7422 & 17 30 37.6 & +74 22 23.8 & 3.20 & wp701200    & Galaxy  &  N S C  & NGC6414, $z$=0.054&\\
RX J1845.6+7956  & 18 45 41.3 &  +79 56 34.5 & 2.23 & rp700058    & Blend   &  X D  & Star:HD175938 + ?& \\ 
RX J2109.7--1332 & 21 09 47.8 & --13 32 24.2 & 1.38 & rp201007n00 & Blend &  S C  & QSO + ?& \\  
RX J2202.8--5636 & 22 02 52.9 & --56 36 08.3 & 1.63 & rp200559n00 & Blend   &  X D  & id pending& \\ 
RX J2215.2--2944 & 22 15 16.4 & --29 44 29.2 & 1.95 & rp701390n00 & Blend   &  N  & {\scriptsize QS0: HB89:2212--299, $z$=2.7 + ?} &35\\  
RX J2223.8--0206 & 22 23 48.8 & --02 06 13.0 & 2.19 & rp701018n00 & Blend   &  S  & AGN, $z$=0.0558 + ?& \\ 
RX J2236.0+3358  & 22 36 00.3 & +33 58 24.0 & 3.18  & wp800066    & Group   &  N  & Stef.Quintet, $z$=0.0215 &36\\  
RX J2237.0--1516 & 22 37 00.6 & --15 16 08.0 & 1.68 & wp201723n00 & Cluster  &  S C  & $z$=0.299& \\ 
RX J2258.1+2055 & 22 58  08.4  & +20 55 15.0 & 2.26 & rp201282n00 & Cluster   &  S N C  & Z2255.5, $z$=0.288 & 37\\   
RX J2309.4--2713 & 23 09 27.9 & --27 13 20.1 & 1.19 & rp900323n00 & Blend  &  S C  & AGN, $z$=0.25 + ?& \\ 
RX J2311.4+1035 & 23 11 25.9  & +10 35 06.7 & 3.52 & rp100578n00 & Blend   &  S C  & AGN, $z$=0.127 + ?& \\ 
RX J2314.7+1915 & 23 14 44.0  & +19 15 23.3 & 1.39 & rp800488n00 & Blend   &  S C  & Cluster, $z$=0.28 + Mstar& \\ 
RX J2353.5--1524 & 23 53 31.5 & --15 24 51.2 & 1.18 & wp701501n00 & Blend   &  S C  & QSO + Mstar& \\ \hline
\end{tabular}

\medskip
\begin{minipage}{0.87\linewidth}
{\sc Notes} ---
Count rates (column 4) are quoted in units of 10$^{-2}$ counts s$^{-1}$.
$^{1}$Redshift taken from \cite{Devaucbc}, D91 hereafter).
$^{2}$Confirmation of redshift provided by Piero Rosati (private communication) and \cite{Ebeling99b}.
$^{3}$Redshift taken from D91.
$^{4}$Redshift taken from \cite{stockebc}, S91 hereafter).
$^{5,6}$Additional spectroscopy provided by Ian Del Antonio.
$^{7}$Reshift taken from \cite {abell89b}.
$^{8}$Redshift taken from D91.
$^{9}$Additional redshift information available in \cite{perlmanb}.
$^{10}$Redshift taken from \cite{laubertsb}. %now a Longetti reference
$^{11}$Redshift taken from D91.
$^{12}$Redshift taken from S91.
$^{13}$Redshift taken from \cite{carballob}.
$^{14}$Redshift taken from \cite{huchrab}.
$^{15}$Redshift taken from \cite{slinglendb}.
$^{16}$Redshift taken from \cite{strubleb}.
$^{17}$Redshift taken from S91.
$^{18}$Redshift taken from D91.
$^{19,20}$Redshift taken from S91.
$^{21}$Redshift taken from D91.
$^{22}$Redshift taken from S91.
$^{23}$Redshift taken from \cite{zabludoffb}.
$^{24}$Redshift taken from D91.
$^{25}$Redshift taken from S91.
$^{26}$Redshift taken from \cite{strubleb}
$^{27}$Redshift taken from \cite{fairallb}.
$^{28}$Additional redshift information available in \cite{boyleb}.
$^{29}$Spectroscopy provided by Dave Turnshek and Eric Monier.
$^{30}$Spectroscopy provided by Ian Del Antonio.
$^{31}$Redshift taken from \cite{perlmanb}.
$^{32}$Also known as EXSS 1646.5+8238 (\cite{tucker}).
$^{33}$Also known as ``Cluster B'' (\cite{reimers}).
$^{34,35}$Redshift taken from \cite{HB89b}.
$^{36}$Redshift taken from \cite{hicksonb}.
$^{37}$Redshift taken from S91.
\end{minipage}
\par
\end{center}
\label{XrayBrightSHARC}
\end{sidetable}
}	%end small
  
%\end{document}

\small{

\begin{sidetable}
\caption{Bright SHARC Cluster Catalog}
\begin{center}
\begin{tabular}{llrrrrrrrrrrlcl} \hline
\multicolumn{1}{c}{Source} & \multicolumn{1}{c}{Redshift} & \multicolumn{1}{c}{nH} & \multicolumn{1}{c}{Major} & \multicolumn{1}{c}{Minor} & \multicolumn{1}{c}{Offaxis} &\multicolumn{1}{c}{$cr_W$} &\multicolumn{1}{c}{$cr_T$} & \multicolumn{1}{c}{$\delta cr_T$}  &  \multicolumn{1}{c}{$r_{80}$} & \multicolumn{1}{c}{$f_{80}$} & \multicolumn{1}{c}{$f_{-13}$} & \multicolumn{1}{c}{$L_{44}$} & \multicolumn{1}{c}{T} & \multicolumn{1}{c}{Notes}\\ 
\multicolumn{1}{c}{(1)}&\multicolumn{1}{c}{(2)}&\multicolumn{1}{c}{(3)}&\multicolumn{1}{c}{(4)}&\multicolumn{1}{c}{(5)}&\multicolumn{1}{c}{(6)}&\multicolumn{1}{c}{(7)}&\multicolumn{1}{c}{(8)}&\multicolumn{1}{c}{(9)}&\multicolumn{1}{c}{(10)}&\multicolumn{1}{c}{(11)}&\multicolumn{1}{c}{(12)}&\multicolumn{1}{c}{(13)}&\multicolumn{1}{c}{(14)}&\multicolumn{1}{c}{(15)}\\	\hline
RX J0117.6--2238 & 0.207 & 1.51 & 8.82 & 6.32 & 86.91 & 1.255 & 2.285 & 9.2\% & 18.36 & 0.943 & 2.594 & 0.4951 & 3& A2894\\
RX J0152.7--1357 & 0.83  & 1.42 & 11.4 & 4.99 & 57.22 & 1.716 & 2.440 & 5.6\% & 9.900 & 1.000 & 2.930 & 8.2604 & 9& \\ 
RX J0221.1+1958 & 0.45   & 9.30 & 6.70 & 5.47 & 73.54 & 1.545 & 2.211 & 6.2\% & 12.03 & 0.969 & 3.296 & 2.8661 & 6& \\ 
RX J0223.4--0852 & 0.163 & 3.18 & 8.86 & 4.98 & 43.25 & 1.202 & 3.036 & 7.7\% & 21.09 & 0.994 & 3.706 & 0.4350 & 3& (1)\\ 
RX J0237.9--5224 & 0.133 & 3.08 & 8.56 & 5.53 & 69.94 & 2.324 & 6.107 & 5.3\% & 24.63 & 0.906 & 7.495 & 0.5824 & 3& A3038\\ 
RX J0250.0+1908 & 0.12   & 9.40 & 13.5 & 5.92 & 57.00 & 1.426 & 1.576 & 15.3\% & 26.49 & 0.997 & 2.242 & 0.1443 & 2& \\ 
RX J0256.5+6.00 & 0.36   & 5.33 & 7.40 & 6.82 & 83.87 & 3.614 & 5.692 & 4.9\% & 13.44 & 0.998 & 7.549 & 4.1597 & 7& \\ 
RX J0318.5--0302 & 0.37  & 5.09 & 6.79 & 5.08 & 43.35 & 2.763 & 4.191 & 5.5\% & 12.85 & 1.000 & 5.587 & 3.2819 & 6& (2)\\ 
RX J0321.9--5119 & 0.0696& 2.46 & 6.93 & 5.36 & 73.91 & 6.180 & 17.10 & 2.4\% & 40.23 & 0.916 & 20.72 & 0.4355 & 3& A3120 (3)\\ 
RX J0426.1+1655 & 0.38   & 16.4 & 5.23 & 4.55 & 22.54 & 1.797 & 2.948 & 6.3\% & 12.62 & 1.000 & 5.159 & 3.1969 & 6& \\ 
RX J0454.3--0239 & 0.26  & 5.24 & 10.4 & 5.92 & 85.65 & 1.594 & 2.732 & 8.0\% & 16.01 & 0.914 & 3.564 & 1.0634 & 4& (4)\\ 
RX J0849.1+3731 & 0.230  & 3.07 & 8.42 & 7.47 & 52.64 & 1.295 & 2.052 & 11.3\% & 16.84 & 0.991 & 2.525 & 0.5976 & 3& A708\\ 
RX J1024.3+6805 & 0.201  & 2.13 & 6.33 & 5.51 & 53.89 & 2.602 & 4.787 & 5.8\% & 18.38 & 0.991 & 5.699 & 1.0100 & 4& A981 (5)\\ 
RX J1113.8+4017 & 0.0795 & 1.80 & 12.6 & 4.36 & 88.73 & 1.440 & 5.879 & 6.1\% & 36.39 & 0.913 & 6.696 & 0.1866 & 2& A1203\\ 
RX J1120.1+4318 & 0.60   & 2.15 & 5.33 & 4.88 & 48.43 & 2.117 & 2.728 & 8.1\% & 10.63 & 1.000 & 3.285 & 5.0100 & 7& \\ 
RX J1142.2+1026 & 0.0698 & 3.33 & 6.15 & 5.88 & 40.27 & 1.272 & 4.245 & 10.8\% & 40.04 & 0.927 & 5.182 & 0.1110 & 2& A1356\\ 
RX J1204.0+2807 & 0.167  & 1.69 & 6.16 & 4.56 & 29.50 & 4.492 & 10.60 & 3.2\% & 20.93 & 0.997 & 12.38 & 1.4942 & 5& MS1201.5\\ 
RX J1211.2+3911 & 0.34   & 2.02 & 6.34 & 4.76 & 62.48 & 1.525 & 2.649 & 5.8\% & 13.52 & 0.988 & 3.163 & 1.5948 & 5& MS1208.7\\ 
RX J1222.1+7526 & 0.24   & 2.88 & 5.14 & 4.48 & 31.01 & 1.175 & 1.353 & 20.5\% & 16.28 & 1.000 & 1.630 & 0.4208 & 3& MS1219.9\\ 
RX J1227.4+0849 & 0.0895 & 1.70 & 9.48 & 7.97 & 74.61 & 5.872 & 19.75 & 2.5\% & 33.15 & 0.983 & 22.57 & 0.7876 & 3& A1541\\ 
RX J1232.8+2605 & 0.22   & 1.36 & 10.4 & 5.72 & 73.53 & 1.828 & 3.577 & 8.0\% & 17.47 & 0.932 & 4.136 & 0.8799 & 4& \\ 
RX J1241.5+3250 & 0.39   & 1.28 & 7.24 & 6.35 & 77.06 & 2.386 & 3.996 & 5.7\% & 12.98 & 0.999 & 4.748 & 3.0995 & 6& \\ 
RX J1259.7--3236 & 0.076 & 5.92 & 7.68 & 4.80 & 77.70 & 1.198 & 4.058 & 11.7\% & 37.59 & 0.884 & 5.222 & 0.1328 & 2& (6)\\ 
RX J1308.5+5342 & 0.33   & 1.59 & 8.57 & 6.20 & 44.76 & 1.254 & 1.732 & 10.7\% & 13.62 & 1.000 & 1.978 & 0.9579 & 4& \\ 
RX J1311.2+3228 & 0.245  & 1.08 & 5.71 & 5.19 & 48.66 & 2.539 & 4.647 & 5.7\% & 16.11 & 1.000 & 5.376 & 1.4223 & 4& MS1308.8\\ 
RX J1311.8+3227 & 0.43   & 1.08 & 5.62 & 5.52 & 73.97 & 1.472 & 2.029 & 9.4\% & 12.22 & 0.913 & 2.375 & 1.9237 & 5& \\ 
RX J1334.3+5030 & 0.62   & 1.08 & 8.84 & 5.04 & 70.18 & 1.366 & 1.810 & 8.4\% & 10.76 & 0.990 & 2.091 & 3.4606 & 6& (7)\\ 
RX J1343.7+5538 & 0.0766 & 1.05 & 12.8 & 6.60 & 68.94 & 1.808 & 9.228 & 4.9\% & 37.31 & 0.946 & 10.32 & 0.2668 & 2& A1783\\ 
RX J1406.9+2834 & 0.117  & 1.40 & 6.68 & 5.01 & 28.69 & 1.304 & 3.625 & 7.3\% & 26.93 & 0.987 & 4.085 & 0.2497 & 2& V154\\ 
RX J1416.4+2315 & 0.138  & 2.04 & 13.7 & 4.87 & 42.77 & 4.804 & 11.18 & 4.6\% & 23.86 & 0.989 & 13.33 & 1.1066 & 4& (8)\\ 
RX J1418.5+2510 & 0.29   & 1.78 & 6.19 & 5.35 & 30.37 & 3.788 & 6.549 & 3.8\% & 14.62 & 1.000 & 7.655 & 2.7618 & 6& V159\\ 
RX J1524.6+0957 & 0.078  & 2.88 & 5.53 & 5.14 & 16.07 & 1.646 & 2.236 & 29.8\% & 36.70 & 0.914 & 2.371 & 0.0649 & 1& V170\\ 
RX J1541.1+6626 & 0.245  & 2.90 & 8.15 & 5.70 & 42.96 & 1.517 & 2.356 & 7.7\% & 16.11 & 0.999 & 2.814 & 0.7578 & 3& (9)\\ 
RX J1641.2+8233 & 0.195  & 5.51 & 12.7 & 7.44 & 68.56 & 3.550 & 6.227 & 5.2\% & 18.87 & 0.993 & 8.128 & 1.3550 & 4& V183\\ 
RX J1701.3+6414 & 0.453  & 2.51 & 5.62 & 4.71 & 13.16 & 1.986 & 3.302 & 4.7\% & 11.62 & 0.997 & 3.965 & 3.4935 & 6& V190\\ 
RX J2237.0--1516 & 0.299 & 3.90 & 11.9 & 5.65 & 87.75 & 1.683 & 2.723 & 7.7\% & 14.91 & 0.990 & 3.413 & 1.3525 & 4& \\ 
RX J2258.1+2055 & 0.288  & 4.91 & 5.74 & 4.81 & 51.96 & 2.262 & 4.428 & 6.1\% & 14.76 & 0.997 & 5.694 & 2.0550 & 5& Z2255.5\\ \hline
\end{tabular}
\medskip
\begin{minipage}{1.12\linewidth}
{\sc Notes} --- Count rates (columns 4 \& 5) are quoted 
in units of 10$^{-2}$ counts s$^{-1}$. 
$^{1}$RX J0223.4 ($z$=0.163) was detected in pointing rp800016n00, 
the central target of which was a wide angle radio (WAR) source. The cluster hosting this WAR 
source has a redshift of $z$=0.41 (\cite{Nichol94}). The redshift 
separation of the two clusters is $\delta z\simeq$0.247. 
$^{2}$RX J0318.5 ($z$=0.37) was detected in wp800555n00 which was pointed, accidentally, $\sim$40$^{\circ}$ away in
declination away from  the listed target, A3112, which lies at 03:17:56--44:14:17 (\cite{Ebeling1}). 
$^{3}$RX J0321.9 (A3120, $z$=0.0696) was detected in wp800371n00, the central 
target of which was a \cite{couchb} cluster at $z$=0.49.  The redshift 
separation of the two clusters is $\delta z$=0.42.  
$^{4}$RX J0454.3 ($z$=0.26) was detected in rp800229n00, the central target of which 
was cluster MS0451.6 ($z$=0.55, \cite{GioiaL}). The redshift separation of the
two clusters is $\delta z$=0.29. 
$^{5}$RX J1024.3 (A981, $z$=0.201) was detected in wp800641, the central target of which
was cluster A998 ($z$=0.202, \cite{huchra}). The redshift separation of the
two clusters is $\delta z$=0.001. 
$^{6}$RX J1259.7 ($z$=0.076) was detected in rp800384n00, the central target of which
was cluster A3537 ($cz$=5007 km/s, \cite{abell89}). The redshift separation of the
two clusters is $\delta z$=0.059.
$^{7}$RX J1334.3 ($z$=0.62) was detected in rp800047, the central target of
which was cluster A1758 ($z$=0.2792, \cite{allen}). The redshift separation of the
two clusters is $\delta z=$0.34.
$^{8}$RX J1416.4 ($z$=0.138)  was detected in rp800401a01, the central target of
which was galaxy 4C23.37 ($cz$=154 km/s, \cite{Devauc}). The redshift separation of the
two clusters is $\delta z$=0.137. 
$^{9}$ RX J1541.1 ($z$=0.245) was detected in rp800511n00, the central target of
which was A2125 ($z$=0.2465, \cite{struble}). The redshift separation of the
two clusters is $\delta z$=0.0015.
\end{minipage}
\par
\label{clusters}
\end{center}
\end{sidetable}
%\end{center}
}

%7/2/99 beta=0.67 have to add the notes in by hand (and decreased sig. figs on errors)
%calculated new fluxes with his rc and redshifts and my background apertures by scaling from ratio of count rates
\begin{table*}
\begin{center}
\caption{Comparison of Bright SHARC and V98 Flux Measurements}
\begin{tabular}{|l|l|l|l|l|l|}\hline
\multicolumn{1}{|c|}{Bright SHARC ID.} & \multicolumn{1}{c|}{V98 ID.} &  \multicolumn{1}{c|}{Redshift}  & \multicolumn{1}{c|}{$f_{-13}$} & \multicolumn{1}{c|}{Ratio$^{1}$} & \multicolumn{1}{c|}{Ratio$^{2}$}\\  \hline
RX J0237.9--5224&V28 & 0.1330   & 7.495 &  1.16 & 1.06 \\ % 6.107/5.05767
RX J0849.1+3731 &V62 & 0.2300   & 2.525 &  1.72 & 1.48 \\ % 2.052/1.76240
RX J1204.0+2807 &V112& 0.1670   & 12.38 &  1.21 & 0.91 \\ % 10.60/8.05895
RX J1211.2+3911 &V115& 0.3400   & 3.163 &  1.19 & 0.83 \\ % 2.649/1.84833
RX J1308.5+5342 &V132& 0.3300   & 1.978 &  1.15 & 0.91 \\ % 4.647/3.67687
RX J1406.9+2834 &V154& 0.1170   & 4.085 &  1.58 & 1.22 \\ % 3.625/2.81214
RX J1418.5+2510 &V159& 0.2900   & 7.655 &  1.01 & 0.95 \\ % 6.549/6.22272
RX J1524.6+0957 &V170& 0.0780   & 2.371 &  0.78 & 0.83 \\ % 2.236/2.3873
RX J1641.2+8233 &V183& 0.1950   & 8.128 &  1.01 & 1.04 \\ % 6.227/6.42979
RX J1701.3+6414 &V190& 0.4530   & 3.965 &  1.03 & 0.92 \\ % 3.302/2.96404
RX J2258.1+2055 &V213& 0.2880   & 5.694 &  1.13 & 0.93 \\ \hline% 4.428/3.66312
\multicolumn{4}{|l|}{Average}           &  1.18 & 1.01 \\ \hline
\end{tabular}
\begin{minipage}{0.87\linewidth}
{\sc Notes} ---
$^1$Ratios of the Bright SHARC fluxes 
(column 4) to the \cite{Viklinin1bc} V98)  fluxes. 
$^2$Ratios of the re-calculated Bright SHARC fluxes 
to the V98 fluxes, see \S\ref{luminbias} for details.
\end{minipage}
\par
\label{vik_comp}
\end{center}
\end{table*}

\end{document}